\documentclass[amsmath,amssymb,showpacs,twocolumn,superscriptaddress,prb]{revtex4-2}
\usepackage{graphicx,bm,color,subfigure}
\usepackage[T1]{fontenc}
\usepackage{mathrsfs}
\usepackage{bm}
\usepackage{amsmath}
\usepackage{amssymb}
\usepackage{graphicx}
\usepackage{esint}
\usepackage{multirow}
\usepackage{float}
\usepackage{array}
\usepackage{makecell}
\usepackage{harpoon}
\usepackage{booktabs}
\usepackage{gensymb}
\usepackage{simplewick}
\usepackage{subfigure}
\usepackage{soul}
\usepackage{makecell}
\usepackage{hyperref}
\hypersetup{colorlinks,allcolors=blue}


\renewcommand{\arraystretch}{1.5}

\begin{document}

\title{Charge-4e/6e superconductivity and chiral metal from 3D chiral superconductor}

\author{Chu-Tian Gao}
\affiliation{School of Physics, Beijing Institute of Technology, Beijing 100081, China}
\author{Chen Lu}
\affiliation{School of Physics, Hangzhou Normal University, Hangzhou 311121, China}
\author{Yu-Bo Liu}
\affiliation{Institute of Theoretical Physics, Chinese Academic of Science, Beijing 100080, China}
\author{Zhiming Pan}
\email{panzhiming@xmu.edu.cn}
\affiliation{Department of Physics, Xiamen University, Xiamen 361005, Fujian, China}
\author{Fan Yang}
\email{yangfan\_blg@bit.edu.cn}
\affiliation{School of Physics, Beijing Institute of Technology, Beijing 100081, China}

\begin{abstract}
Unconventional superconductivity (SC) characterized by multi-fermion orderings has attracted substantial attention. 
However, previous studies have largely focused on 2D systems or 3D systems with effective 2D symmetries. 
Here, we investigate the vestigial phases arising from thermal fluctuations of chiral SC in 3D systems governed by the cubic $O_h$ point group. 
By constructing low-energy effective Hamiltonians via Ginzburg-Landau analysis and conducting Monte Carlo simulations, 
we systematically investigate the phase fluctuations of chiral orders within the $E_g$ and $T_{2g}/T_{1u}$ irreducible representations (IRRPs). 
We identify a phase diagram topology different from 2D counterparts,
where the multi-phase intersection manifests as a tetracritical point rather than the triple point typically found in 2D systems. 
We elucidate the evolution of these phases under thermal fluctuations.
Our findings reveal that for both $E_g$ and $T_{2g}/T_{1u}$ IRRPs, the primary chiral orders could melt into a chiral metallic phase across specific parameter regimes.
Moreover, for the $E_g$ IRRP, phase fluctuation could also induce a charge-$4e$ phase under certain regime, while for the $T_{2g}$ and $T_{1u}$ IRRPs, it leads to a higher-order charge-$6e$ SC state.
Our work paves the way for exploring exotic vestigial orders driven by non-trivial 3D crystalline symmetries.
\end{abstract}

\maketitle
\section{Introduction}
The study of vestigial phases arising from fluctuating superconductivity (SC) has evolved into an important research frontier~\cite{korshunov1985,kivelson1990doped,ropke1998four,douccot2002pairing,moore2004,babaev2004phase,wu2005competing,aligia2005quartet,agterberg2008dis,berg2009charge4e,agterberg2011con,ko2009doped,herland2010phase,you2012super,jiang2017charge4e,li2017nematic,yonezawa2017therm,tao2018direct,kostylev2020uniaxial,le2020evidence,jian2021charge4e,fu2021charge4e,grinenko2021state,han2022,lothman2022nematic,zhou2022chern,song2022phase,rampp2022topo,varma2023extended,yu2023non,curtis2023stabliz,hecker2023cascade,liu2023charge4e,li2024charge4e,zhang2024higgs,poduval2024vestigial,zeng2024high,hecker2024local,wu2024dwave,liu2024nematic,volovik2024fermionic,ge2024charge4e,pan2024frustrated,lin2025theory}. 
These exotic states can emerge when strong thermal or quantum fluctuations melt the conventional charge-$2e$ SC order while preserving higher-order composite correlations.
Although a primary SC phase typically breaks multiple symmetries simultaneously (e.g., global gauge, time-reversal and/or crystalline symmetries), fluctuations can destroy the phase coherence of individual Cooper pairs. 
Simultaneously, composite degrees of freedom can still remain phase-coherent, leading to partial symmetry breaking.
This may establish a residual long-range order (or quasi-long-range order in 2D), stabilizing intermediate vestigial phases such as charge-$4e$/$6e$ SC, chiral metals, or nematic orders.

%The physical mechanisms driving the formation of these vestigial phases depend on the dimensionality of the system. 
In 2D systems, the melting of the quasi-long-ranged primary order is typically governed by the Berezinskii-Kosterlitz-Thouless (BKT) mechanism due to the absence of true long-range order.
Vestigial orders are realized through the proliferation and unbinding of topological excitations.
The selective unbinding of different vortex species restores the symmetries, 
naturally opening intermediate temperature windows for vestigial phases. 
In contrast, the physics of their 3D counterparts with long-range order is different. 
The 3D phase transitions are primarily governed by the conventional thermal fluctuations of the order parameters rather than vortex unbinding. 
Previous studies on 3D nematic superconductors have already indicated the existence of charge-$4e$ SC and nematic order driven by such 3D fluctuations ~\cite{jian2021charge4e,fu2021charge4e}.

Recently, the experimental realization of the 3D Hubbard model using optical lattices in cold atom systems has attracted significant interest ~\cite{morong2021direct,iskakov2022phase,lenihan2022evaluating,shao2024afm,langmann2025univer,song2025extended,song2025magnetic,sun2025boosting}. 
Microscopic analysis suggest that SC could emerge in these systems upon doping, accompanying the suppression of the antiferromagnetic order ~\cite{raugh2010sc,pan2024octupolar}. 
Governed by the cubic crystal symmetry (point group $O_h$),
the pairing symmetry could be classified by irreducible representations (IRRPs) ~\cite{sigrist1991}.
The dominant pairing channel is predicted to evolve with doping: transitioning from the $E_g$ IRRP at low-to-moderate doping to the $T_{2g}$ (or $T_{1u}$) IRRP at higher doping levels ~\cite{pan2024octupolar}.
Here, $E_g$ state serves as the 3D counterpart to the well-known $d_{x^2-y^2}$-wave pairing in 2D square lattice near half-filling.
All these relevant IRRPs ($E_g$,$T_{2g}$,$T_{1u}$) are multi-dimensional IRRPs.
Ginzburg-Landau (G-L) analysis further demonstrate that the ground states of such multi-component order parameters typically break additional symmetries spontaneously, 
realizing either a chiral SC state which breaking time-reversal symmetry (TRS) or a nematic state which breaks the full rotational symmetry of the lattice ~\cite{sigrist1991,pan2024octupolar}.

In this paper, we investigate the low-energy behavior of fluctuating SC through symmetry analysis and G-L free energy theory. 
Specifically, we focus on chiral phases governed by the IRRPs of the 3D cubic $O_h$ point group.
These include the two-component $E_g$ IRRP with the order parameter $\Delta= (\Delta_1, \Delta_2)$ and the three-component $T_{2g}/T_{1u}$ IRRPs with $\Delta= (\Delta_1, \Delta_2,\Delta_3)$. 
In the ground state, these components form chiral superpositions, such as $\Delta_1+i\Delta_2$ for $E_g$ case, 
or $\Delta_1+e^{i2\pi/3} \Delta_2+e^{i4\pi/3}\Delta_3$ for $T_{2g}/T_{1u}$ cases, 
that spontaneously breaks both TRS and global $U(1)$ symmetry. 
At finite temperatures, thermal fluctuations can partially melt this primary superconducting order, thereby stabilizing exotic vestigial phases.
If only TRS is restored while the global $U(1)$ phase coherence is preserved, the system enters a charge-$4e$ or $6e$ SC phase, which can be phenomenologically characterized by composite order parameters such as $\Delta_1 \Delta_2$ or $\Delta_1 \Delta_2\Delta_3$, respectively.
Conversely, if the global $U(1)$ gauge symmetry is restored but TRS remains broken, the system transitions into a chiral metal state, driven by the relative phase order $\Delta_a \Delta^{\ast}_b (a \neq b)$. 
Different from 2D systems, which typically exhibit a triple point in their phase diagrams due to vortex-mediated BKT transition, 
3D phase diagrams universally feature a tetracritical point.
Furthermore, we demonstrate that the $T_{2g}/T_{1u}$ IRRPs uniquely support a higher-order charge-$6e$ SC state, highlighting the richness of vestigial orders in systems with full 3D crystalline symmetries.

The rest of this paper is organized as follows. In Sec.~\ref{sec:model}, we have derived low-energy effective Hamiltonians for the chiral superconducting state formed by the $E_g$, $T_{2g}/T_{1u}$ order parameter using the G-L free energy analysis.
%In Sec.~\ref{sec:Eg_nematic}, we present the MC results for the nematic superconducting state formed by the $E_g$ order parameter. 
In Sec.~\ref{sec:Results and discussion}, we present and discuss the Monte Carlo (MC) results for these different representations, including phase diagrams, transition properties, and correlation functions. Finally, in Sec.~\ref{sec:CON}, we provide a conclusion.

\section{Ginzburg-Landau analysis}
\label{sec:model}

\begin{figure}[t!]
\centering
\includegraphics[width=0.45\linewidth]{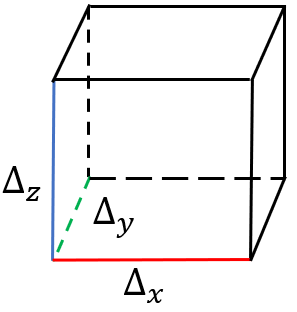}
\caption{(Color online) Superconducting order parameter on a cubic lattice.}\label{cubic_SC}
\end{figure}

We consider a 3D cubic lattice system governed by the $O_h$ symmetry.
To capture the lattice geometry, we define nearest-neighbor pairing amplitudes along the three principal axes, denoted as $(\Delta_x,\Delta_y,\Delta_z)$, as schematically illustrated in Fig.~\ref{cubic_SC}. 
These bond parameters can be decomposed according to the IRRPs of the symmetry group. 
The isotropic configuration, $(1,1,1)$, corresponds to the trivial $A_{1g}$ ($s$-wave) pairing symmetry. 
In contrast, the linearly independent anisotropic combinations, proportional to $ (1,-1,0)$ and $(-1,-1,2)$, transform as the $d_{x^2-y^2}$ and $d_{3z^2-r^2}$ orbital symmetries, respectively.
These two basis functions span the two-dimensional $E_g$ IRRP.
We note that realizing other IRRPs, such as $T_{2g}$ or $T_{1u}$, would require incorporating next-nearest-neighbor or longer-range pairings.

\subsection{Chiral SC within $E_g$ IRRP}
Here, we specifically investigate the chiral SC state associated with the $E_g$ IRRP ~\cite{sigrist1991,pan2024octupolar}.
This phase is characterized by a two-component complex order parameter,
$\psi=(\psi_1,\psi_2)$, where the components correspond to the amplitudes of the two distinct $d$-wave basis functions:
$\Delta_1\equiv \Delta_{d_{x^2-y^2}}$ and $\Delta_2\equiv\Delta_{d_{3z^2-r^2}}$. 
The total gap function is expressed as a superposition:
\begin{equation}
\Delta=\psi_{1}\Delta_{1} + \psi_{2}\Delta_{2}.
\end{equation}
To determine the ground state structure, we employ the the G-L free energy analysis (see appendix for details).
The free energy functional $F$ must respect the symmetries of the system, 
including the $O_{h}$ point group, global $U(1)$ symmetry and TRS. 
Truncating the expansion at the fourth order, the free energy takes the form:
\begin{equation}
F= \alpha ( |\psi_{1}|^2 +|\psi_{2}|^2 ) 
+F^{(4)} +\cdots,
\label{eq:EgFreeEnergy}
\end{equation}
where the quartic term respecting the symmetries is given by ~\cite{sigrist1991},
\begin{equation}
\begin{aligned}
F^{(4)} =& \beta_1 ( |\psi_{1}|^{2}+|\psi_{2}|^{2} )^2 
+\beta_2 \big|\psi^{\ast}_{1} \psi_{2} -\psi^{\ast}_{2} \psi_{1}\big|^2.
\end{aligned}
\end{equation}
Here, $\alpha$, $\beta_1$ and $\beta_2$ are the phenomenological G-L parameters.
In the ordered phase where $\alpha<0$ and $\beta_1>0$, the symmetry pattern is dictated by the sign of $\beta_2$-term.
Specifically, $\beta_2<0$ stabilizes a complex mixing of the order parameters with a relative phase of $\pm\pi/2$, i.e., $(1:i)$ or $(i:1)$, 
leading to a chiral SC state that spontaneously breaks TRS. 
Conversely, for $\beta_2>0$, a nematic pairing state with real mixing is stabilized.
In the following discussion, we focus on the chiral case which is achieved below the transition temperature $T_c$.

Slightly above the transition temperature $T_c$, the system's behavior is dominated by phase fluctuations of the SC order parameter.
We parameterize the pairing order parameter by setting $\psi_i=\psi_{0}e^{i\theta_i(\bm{r})}$, assuming a constant amplitude $\psi_{0}>0$. 
It is convenient to decouple the fluctuations into a total phase $\theta(\bm{r})$ and a relative phase $\phi(\bm{r})$,
defined such that: $\theta_1=\theta+\phi$ and $\theta_2=\theta-\phi$.
Substituting this parameterization into the G-L free energy and including terms involving phase gradients ($\nabla\theta$ and $\nabla\phi$), 
the effective low-energy Hamiltonian is given by:
\begin{eqnarray}\label{Hamiltonian_r_Eg}
H = \int d^{3}\bm{r}\Big(\frac{\rho}{2} | \nabla \theta|^{2} + \frac{\kappa}{2} |\nabla \phi|^{2}  + A \cos(4\phi )\Big).
\end{eqnarray}
Here $\rho$ and $\kappa$ represents the stiffness parameters, and in general, $\rho\neq \kappa$ .
The coupling constant for the $Z_4$ anisotropy is $A=-2\beta_2\psi^{4}_{0}$.

It is important to address the potential cross-gradient couplings between the total and relative phases, $\theta$ and $\phi$. 
A globally isotropic cross-term, $\nabla\theta \cdot \nabla\phi \propto |\nabla\theta_1|^2 - |\nabla\theta_2|^2$, transforms as the two-dimensional $E_g$ IRRP. 
It cannot independently form a totally symmetric scalar and is forbidden by the $O_h$ point group requirement for the free energy. 
Nevertheless, spatially anisotropic cross-couplings are symmetrically allowed. 
For example, specific combinations of momenta, such as $2k_z^2 - k_x^2 - k_y^2$, also transform as $E_g$, and they can couple with the $E_g$ phase combinations to generate a valid $A_{1g}$ invariant. 
This would induce the cross-gradient coupling between the two phases.
Here we focus on the simpler situation where two phases are nearly decoupled in the low-energy theory.

The effective model Eq.~(\ref{Hamiltonian_r_Eg}) suggests that low-energy properties of the $\theta$ field are similar to those of a 3D classical XY model~\cite{janke1990}.
Meanwhile, the $\phi$ field is described by a specific 3D XY model subject to a $Z_4$ anisotropy term, which reduced to $Z_2$ symmetry physically. 
It is important to note that the states $(\theta(\bm{r}),\phi(\bm{r})+\pi)$ and $(\theta(\bm{r}),\phi(\bm{r}))$ are gauge equivalent, 
as the corresponding physical configurations $\theta_{1,2}\left(\bm{r}\right)$ differ only by a global constant $\pi$~\cite{liu2023charge4e,liu2024nematic}. 
Consequently, the phase transition associated with the $\theta$-field falls into the 3D XY universality class, 
whereas the transition associated with the $\phi$-field belongs to the 3D Ising-type university class ~\cite{hasenbusch2019monte}.

The emergence of vestigial phases is governed by the interplay between phase fluctuations in the global ($\theta$) and relative ($\phi$) phase channels.
At zero temperature, the ground state corresponds to a chiral SC state that spontaneously breaks both TRS and the global $U(1)$ symmetry.
At finite temperatures, however, the stability of these orders is dictated by their respective phase stiffnesses. 
While the presence of a finite anisotropy term $A$ may renormalize the effective relative stiffness $\kappa$ and modify the critical temperature, the qualitative features of the phase diagram remain robust. 
Therefore, the phase boundaries are primarily determined by the competition between $\rho$ and $\kappa$.

First, in the regime where relative phase fluctuations dominate ($\kappa\ll \rho$), 
the relative phase field disorders first upon heating.
This transition restores the discrete $\mathbb{Z}_2$ (Ising) symmetry associated with TRS, while the global $U(1)$ of the superconductor remains broken.
Consequently, although the primary order parameters vanish $\langle\psi_{1,2}\rangle=0$, 
the system retains a composite order parameter $\Delta_{4e}\sim \psi_1 \psi_2\propto e^{2i\theta}$.
This state defines a vestigial charge-$4e$ SC phase, which preserves TRS but exhibits fractionalized flux quantization ($hc/4e$).

Conversely, when the relative phase stiffness is dominant ($\kappa\gg \rho$), the global SC coherence is destroyed first, whereas the relative phase correlations persist.
In this scenario, the $U(1)$ symmetry is restored, destroying SC, but the system remains in a state with broken TRS. 
This results in a chiral metal phase, characterized by the composite order parameter $i(\psi_1\psi_2^*-\psi_1^*\psi_2)\propto \sin(2\phi)$.

Finally, in the intermediate regime where the stiffnesses are comparable ($\kappa\sim\rho$), the global and relative phases may disorder simultaneously. 
In contrast, the 3D system lacks the BKT transition mechanism; the phase transitions are instead driven by conventional thermal fluctuations of the 3D order parameters (belonging to the 3D XY or Ising universality classes). 
Consequently, the 3D fluctuations stabilizes a single tetracritical point, where four distinct phases, the chiral SC, the vestigial charge-$4e/6e$ SC, the chiral metal, and the normal state, converge.

\subsection{Chiral SC within $T_{2g}$/$T_{1u}$ IRRP}
Next, we extend our analysis to  chiral SC state associated with the three-dimensional $T_{2g}$ and $T_{1u}$ IRRPs. 
The basis gap functions $(\Delta_1,\Delta_2,\Delta_3)$ for these IRRPs correspond to the \{$d_{xy}$, $d_{yz}$, $d_{zy}$\} and \{$p_x$, $p_y$, $p_z$\} channels, respectively. 
The total gap function is expressed as a superposition: 
\begin{equation}
\Delta=\psi_{1}\Delta_{1} + \psi_{2}\Delta_{2}+ \psi_{3}\Delta_{3},
\end{equation}
with the pairing coefficients $\psi_a$ ($a=1,2,3$).
The possible configurations are summarized in Tab.~\ref{tab:T2gGroundState}.

The free energy $F$ shares the same form for both representations, as detailed in the appendix. 
Expanding $F$ to the fourth order in the order parameters yields ~\cite{sigrist1991}:
\begin{equation}
\begin{aligned}
&F^{(2)}+F^{(4)} =\alpha \sum_a |\psi_a|^2 
+(\beta_1+\beta_2) \sum_a |\psi_{a}|^{4}    \\
&\qquad +2\beta_2 \sum_{a< b} \psi^{\ast2}_{a}\psi_{b}^{2}
+ 2(\beta_1+\beta_3) \sum_{a<b} |\psi_{a}|^2|\psi_{b}|^{2},
\end{aligned}
\label{eq:TgFreeEnergy}
\end{equation}
with $a,b=1,2,3$. Here, $\alpha$, $\beta_1$, $\beta_2$ and $\beta_3$ are the phenomenological G-L parameters.
In the ordered phase where $\alpha<0$, assuming $\beta_{1}>0$ sufficiently positive to ensure global stability, the ground state symmetry is determined by the competition between $\beta_2$ and $\beta_3$, as summarized in Tab.~\ref{tab:T2gGroundState}.
We explicitly focus on the regime where $\beta_2>0>\beta_3$, which could stabilize a complex superposition of the order parameters with equal amplitudes and relative phase differences of $\pm\frac{2\pi}{3}$, 
denoted as $(1:e^{i\frac{2\pi}{3}}:e^{i\frac{4\pi}{3}})$.
Such specific configuration realizes a chiral SC state that spontaneously breaks TRS. 
There also exists other possibility which spantaneoulsy TRS as Tab.~\ref{tab:T2gGroundState}, i.e., $(1:i:0)$, while such configuration further breaks the $O_h$ rotational symmetry of the SC amplitude.

\begin{table}[t!]
\centering
\begin{tabular}{|c|c|c|}
\hline
$(\psi_1,\psi_2,\psi_1)$  & Free energy & condition    \\
\hline
$|\psi|e^{i\theta} (1,0,0)$ &  $-\alpha^2/4(\beta_1+\beta_2)$ & $\beta_3>0,2\beta_2<\beta_3$ \\
\hline
$|\psi|e^{i\theta} (1,1,0)$ &  $-\alpha^2/4(\beta_1+\beta_2+\beta_3/2)$ & \# \\
\hline
$|\psi|e^{i\theta} (1,i,0)$ &  $-\alpha^2/4(\beta_1+\beta_3/2)$ & $0<\beta_3<2\beta_2$  \\
\hline
$|\psi|e^{i\theta} (1,1,1)$ &  $-\alpha^2/4(\beta_1+\beta_2+2\beta_3/3)$ & $\beta_2,\beta_3<0$  \\
\hline
$|\psi|e^{i\theta} (1,\omega,\omega^2)$ &  $-\alpha^2/4(\beta_1+2\beta_3/3)$ & $\beta_3<0<\beta_2$  \\
\hline
\end{tabular}
\caption{Possible ground state configurations $(\Delta_1, \Delta_2, \Delta_3)$ and their associated free energy. 
The conditions for each configuration (assuming $\beta_1$ is positive large enough) are listed in the third column.
Here, $\omega=e^{i2\pi/3}$.}
\label{tab:T2gGroundState}
\end{table}

To investigate the phase fluctuations as temperature increasing, 
we parameterize the order parameters as $\psi_a=\psi_{0}e^{i\theta_a(\bm{r})}$ with a constant amplitude $\psi_0$. 
We can further decompose the phases into a global phase $\theta(\bm{r})$ and relative phases $\phi_{a}(\bm{r})$ via $\theta_a(\bm{r})=\theta(\bm{r})+\phi_{a}(\bm{r})$ for $a=1,2,3$.
Note that the relative phases satisfy the constraint $\sum_{a}\phi_a=0$, ensuring there only three independent phase degrees of freedom.
Substituting this parameterization into the free energy and retaining the relevant gradient terms ($\nabla \theta$ and $\nabla \phi$) leads to the following low-energy effective Hamiltonian:
\begin{equation}
\begin{aligned}
H =& \int d^{3}\bm{r} \Big( \frac{\rho}{2} | \nabla \theta|^{2} + \frac{\kappa}{2} \sum_{a}|\nabla \phi_{a}|^{2} \Big) +H_A,
\end{aligned}
\label{Hamiltonian_r_T2g}
\end{equation}
subject to the anisotropic term among the relative phases 
\begin{align*}
H_A = A \int d^{3}\bm{r}  \sum_{a<b} \cos(2\phi_a-2\phi_b) .
\end{align*}
Here, the coupling constant is $A=2\beta_2\psi^{4}_{0}$. 
Physically, the total phase field $\theta$ (governed by stiffness $\rho$) falls into the standard 3D XY university class~\cite{janke1990}.
In contrast, considering $3\sum_{a}|\nabla \phi_{a}|^{2}=|\nabla (\phi_1-\phi_2)|^2+ |\nabla(\phi_2-\phi_3)|^2+ |\nabla (\phi_3-\phi_1)|^2$, the relative phase fields $\phi_a-\phi_b$ (associated with stiffness $\kappa$) are described by a 3D XY model modified by a symmetry-breaking $Z_2$ anisotropy term.

The phase diagram for the $T_{2g}/T_{1u}$ IRRPs is also governed by the interplay between global ($\rho$) and relative ($\kappa$) phase stiffnesses, analogous to the $E_g$ case.
In the regime where relative phase fluctuations dominate ($\kappa\ll \rho$), 
the relative order melts prior to the global phase, restoring time-reversal symmetry while preserving superconducting coherence. 
Crucially, the three-component nature of the order parameter stabilizes a unique vestigial charge-$6e$ SC phase $\Delta_{6e}\sim\psi_1\psi_2\psi_3$ distinct from the charge-4e state in the $E_g$ case and characterized by fractionalized flux quantization ($hc/6e$).
Conversely, when the global phase stiffness is weaker ($\kappa\gg \rho$), SC is destroyed while chiral correlations persist, resulting in a chiral metal phase described by the composite order parameter ($i(\psi_a\psi_b^*-\psi_a^*\psi_b)$ with $a\neq b$). 
Between these two asymptotic limits, the critical boundaries for the global and relative phase transitions intersect directly under specific balance of the stiffnesses. 
At this crossing point, the simultaneous disordering of both sectors establishes a tetracritical point, where all four phases converge.

\section{Numerical Phase Diagram} 
\label{sec:Results and discussion}

In this section, we present the comprehensive numerical results obtained from large-scale MC simulations based on the effective models derived in Sec.~\ref{sec:model}. 
Our primary objective is to first discrete the continuum model into lattice one, 
and then determine the finite-temperature phase diagrams numerically.
We will elucidate the nature of the phase transitions, which are driven by the interplay between the global phase stiffness $\rho$ and the relative phase stiffness $\kappa$. 

The discussion is organized by first examining the two-component $E_g$ representation, followed by an analysis of the three-component $T_{2g}/T_{1u}$ representations.

\subsection{$E_g$ representation}
We begin our analysis with the chiral SC state associated with the $E_g$ IRRP.
To enable MC simulations, we discretize the low-energy continuum Hamiltonian Eq.~(\ref{Hamiltonian_r_Eg}) onto a 3D cubic lattice.
It is important to emphasize that while the specific choice of lattice regularization is not unique, the universal critical properties of the phase transitions are robust and independent of such microscopic details;
they are governed by the same low-energy effective field theory.

We construct the lattice model using the standard XY-type formulation.
This approach replaces the continuous spatial gradient of a phase field, $\nabla\theta(\bm{r})$, with a discrete phase difference across nearest-neighbor sites $i$ and $j$, denoted as $\Delta\theta_{ij}\equiv \theta(\bm{r}_i) -\theta(\bm{r}_j)$ (setting the lattice constant $a=1$).
Furthermore, because the physical SC phases are periodic modulo $2\pi$, we naturally replace the unbounded quadratic gradient terms with periodic cosine functions. 
In the long-wavelength limit (the slow-mode approximation), the phase variation between adjacent sites is sufficiently small ($\Delta\theta_{ij}\ll 1$), allowing us to utilize the Taylor expansion $1-\cos(\Delta\theta_{ij})\approx \frac{1}{2}(\Delta\theta_{ij})^2$.
By dropping the constant energy offset, the quadratic gradient $|\nabla\theta|^2$ is mapped to the discrete form $\propto -\sum_{\mu}\cos(\Delta\theta_{i,i+\hat{\mu}})$. Here, the summation index $\mu$ runs over the three nearest-neighbor spatial directions ($\mu=x,y,z$) on the cubic lattice.

For the $E_g$ chiral state, the system is described by two SC phase components, $\theta_1$ and $\theta_2$.
To capture the underlying physics, we group these into a global phase channel $(\theta_1+\theta_2=2\theta)$ and a relative phase channel $(\theta_1-\theta_2=2\phi)$, as encoded in the low-energy Hamiltonian Eq.~(\ref{Hamiltonian_r_Eg}).
In MC simulations, it is crucial to faithfully respect the topological structure of the phase space. 
While the composite variables $\theta$ and $\phi$ are subject to coupled periodicity, the original variables $\theta_1$ and $\theta_2$ are conventional $U(1)$ phases that independently obey a simple modulo $2\pi$ periodicity. 
Therefore, to ensure that this periodic nature is unambiguously captured without introducing artificial gauge constraints, we treat $\theta_1$ and $\theta_2$ directly as the independent degrees of freedom to be sampled.

Applying the previous discretization scheme to the continuum Hamiltonian Eq.~(\ref{Hamiltonian_r_Eg}), the resulting Hamiltonian on the cubic lattice is given by the following nearest-neighbor cosine interactions:
\begin{eqnarray}\label{Hamiltonian_d_Eg}
H &=& -\frac{\rho}{4}\sum_{\langle ij\rangle} \cos(\theta_{1}(\bm{r}_{i})+\theta_{2}(\bm{r}_{i})-\theta_{1}(\bm{r}_{j})-\theta_{2}(\bm{r}_{j}))  \nonumber\\
&&- \frac{\kappa}{4} \sum_{\langle ij\rangle} \cos(\theta_{1}(\bm{r}_{i})-\theta_{2}(\bm{r}_{i})-\theta_{1}(\bm{r}_{j})+\theta_{2}(\bm{r}_{j})) \nonumber\\
&&+ A\sum_{i} \cos(2\theta_{1}(\bm{r}_{i})-2\theta_{2}(\bm{r}_{i})).
\end{eqnarray}
Here, the summation $\langle ij\rangle$ runs over all the nearest-neighbor bonds.
The coupling coefficients $\rho, \kappa$ are positive, ensuring that the model recovers Eq.~(\ref{Hamiltonian_r_Eg}) in the long-wavelength limit.

To explicitly demonstrate how this discrete lattice model reduces to the continuum effective theory Eq.~(\ref{Hamiltonian_r_Eg}) in the long-wavelength limit, we can express the arguments of the cosine functions in terms of the global phase $\theta$ and the relative phase $\phi$.
Recalling the definitions $2\theta=\theta_1+\theta_2$ and $2\phi=\theta_1-\theta_2$, 
the phase differences across a nearest-neighbor bond $\langle ij\rangle$ can be rewritten as $\theta_{1,i}+\theta_{2,i}-\theta_{1,j}-\theta_{2,j}=2\Delta\theta_{ij}$ and $\theta_{1,i}-\theta_{2,i}-\theta_{1,j}+\theta_{2,j}=2\Delta\phi_{ij}$.
Substituting these into the first two terms of Eq.~(\ref{Hamiltonian_d_Eg}) and applying the slow-mode Taylor expansion $\cos(x)\approx 1-x^2/2$, we obtain
\begin{align*}
-\frac{\rho}{4} \cos(2\Delta\theta_{ij}) 
&\approx -\frac{\rho}{4} + \frac{\rho}{2}(\Delta\theta_{ij})^2, \\
-\frac{\kappa}{4} \cos(2\Delta\phi_{ij}) 
&\approx -\frac{\kappa}{4} + \frac{\kappa}{2}(\Delta\phi_{ij})^2.
\end{align*}
By dropping the irrelevant constant energy offsets ($-\rho/4$ and $-\kappa/4$), the kinetic terms yield $\frac{\rho}{2}(\Delta\theta_{ij})^2$ and $\frac{\kappa}{2}(\Delta\phi_{ij})^2$.
Simultaneously, the argument in the local anisotropy term simply evaluates to $2\theta_{1,i}-2\theta_{2,i}=4\phi_i$, which directly yields the $A\cos(4\phi)$ potential. 
Finally, by taking the continuum limit, 
where discrete differences are replaced by spatial gradients ($\Delta\theta_{ij} \rightarrow \nabla\theta$ and $\Delta\phi_{ij} \rightarrow \nabla\phi$) and the lattice summation transitions into a spatial integral ($\sum\rightarrow \int d^3\bm{r}$), we recover the continuum low-energy Hamiltonian presented in Eq.~(\ref{Hamiltonian_r_Eg}).

With the lattice Hamiltonian established above, we proceed to the MC simulations to determine the phase diagrams and critical properties. 
For the numerical implementation, we fix the anisotropy coupling strength at a representative small value of $A=0.05\rho$ for the $E_g$ Eq.~(\ref{Hamiltonian_d_Eg}). 
It is important to emphasize that the exact magnitude of this small anisotropy does not qualitatively alter the resulting phase diagram. 
Specifically, the topological structure of the phase diagram, including the sequence of the vestigial phases and the emergence of the multicritical point, remains robust as long as $A$ is finite and small.
This choice allows us to investigate the effects of symmetry-breaking anisotropy within the critical region. 
Detailed definitions and formulas for the calculated observables are provided in the appendix.

\begin{figure}[t]
\centering
\includegraphics[width=0.3\textwidth]{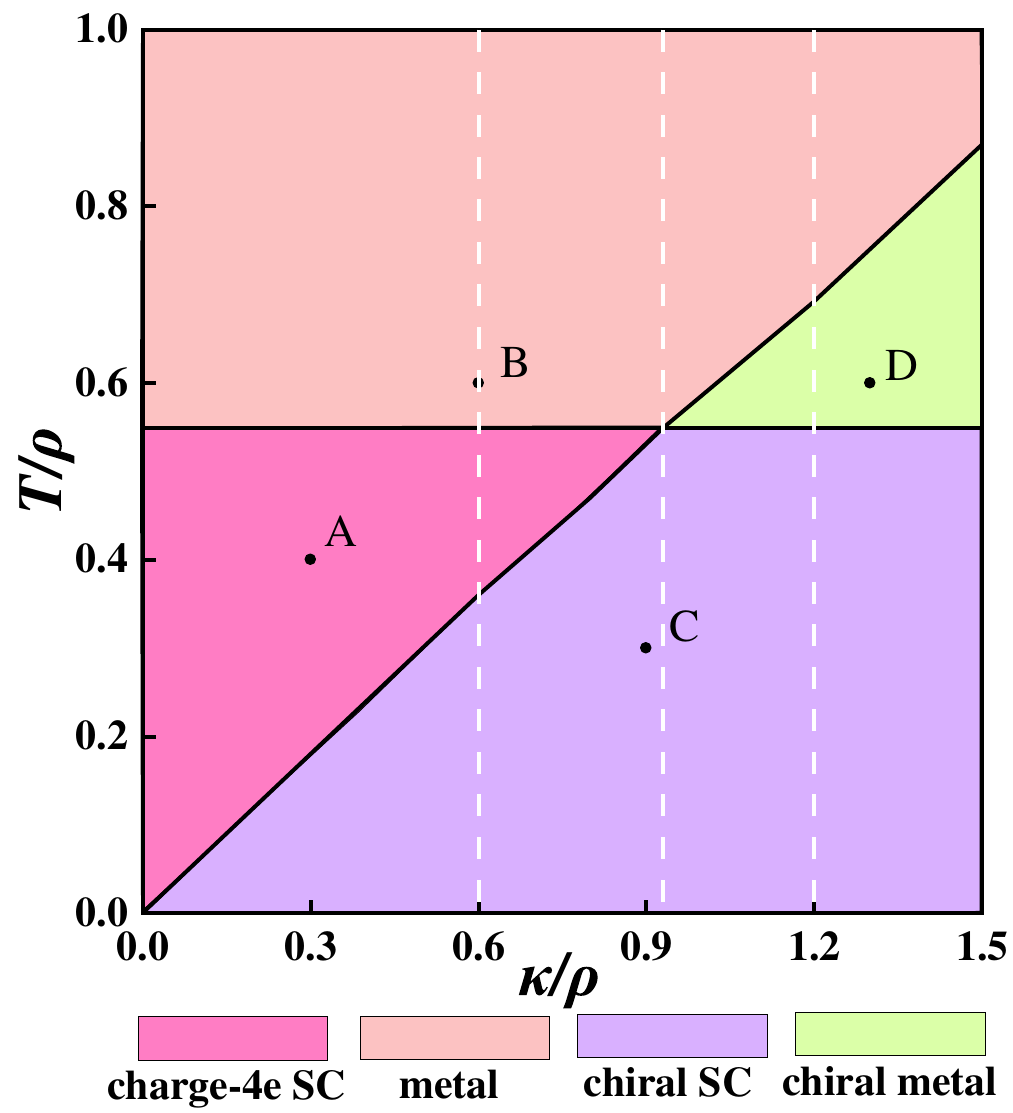}
\caption{(Color online) Phase diagram provided by the MC study for the chiral SC with the order parameters in the $E_g$ representation with $A=0.05\rho$ in Eq. (\ref{Hamiltonian_d_Eg}). The white dashed lines mark $\kappa/\rho=0.6, 0.93$ and $1.2$, respectively. }
\label{Eg_chiral}
\end{figure}

\begin{figure}[t]
\centering
\includegraphics[width=0.44\textwidth]{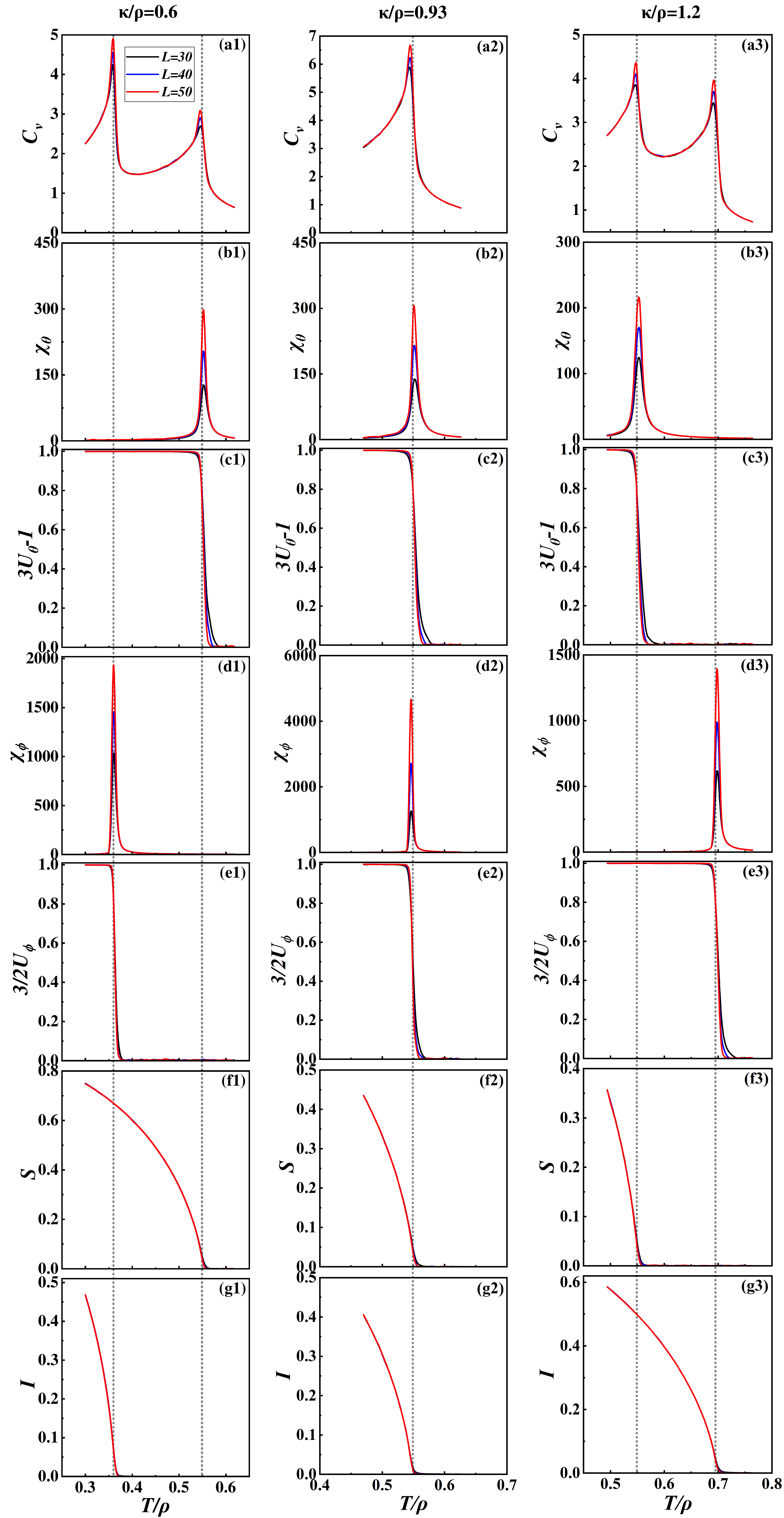}
\caption{(Color online) The quantities as functions of temperature for $\kappa/\rho=0.6$ (a1,b1,...,g1), $\kappa/\rho=0.93$ (a2,b2,...,g2) and $\kappa/\rho=1.2$ (a3,b3,...,g3) (the $E_{g}$ representation). The scaling in all figures is $L=$ 30 (black line), 40 (blue line), and 50 (red line). (a1-a3) The specific heat $C_v$. (b1-b3) The susceptibility $\chi_{\theta}$ of $\theta$ field. (c1-c3) $3U_{\theta}-1$, where $U_{\theta}$ is the Binder cumulant of the $\theta$-field. (d1-d3) The susceptibility $\chi_{\phi}$ of $\phi$ field. (e1-e3) $\frac{3}{2}U_{\phi}$, where $U_{\phi}$ is the Binder cumulant of the $\phi$-field. (f1-f3) The phase stiffness $S$ of $\theta$ field. (g1-g3) The Ising order parameter $I$ of $\phi$ field. The grey dashed lines represent the phase transitions in (a1)-(g3).}
\label{Eg_chiral_ob}
\end{figure}

The numerical results are summarized in the phase diagram depicted in Fig.~\ref{Eg_chiral}. 
To characterize the phase transitions, we numerically simulate the specific heat $C_{v}$, the susceptibility $\chi_{\theta}$ and $\chi_\phi$, the Binder cumulant $3U_{\theta}-1$ of the $\theta$-field, the Binder cumulant $\frac{3}{2}U_{\phi}$ of the $\phi$-field,
the phase stiffness $S$ with $\theta$, and the Ising order parameter $I$ associated with $\phi$.
Fig.~\ref{Eg_chiral_ob} displays the temperature dependence of the states for lattice sizes $L=30, 40, 50$ at three representative coupling ratios: $\kappa/\rho=0.6,0.93$ and $1.2$, corresponding to the vertical cuts marked in Fig.~\ref{Eg_chiral}. 
Additional data and detailed definitions of these observables are provided in the appendix.

Based on the analysis of thermodynamic observables for different values of $\kappa/\rho$, we identify distinct phase transitions: 
(i). For $\kappa/\rho=0.6$, the numerical results are summarized in Fig.~\ref{Eg_chiral_ob}(a1,b1,...,g1).
As the temperature $T/\rho$ increases, the system undergoes two successive transitions.
At $T/\rho\approx 0.36$, the specific heat $C_v$ exhibits a peak that scales with system size $L$. 
Concurrently, for $\phi$-field, the susceptibility $\chi_{\phi}$ diverges, 
while the cumulant $\frac{3}{2}U_{\phi}$ and the Ising order parameter $I$ drops to zero. 
This indicates that the $\phi$-field undergoes an Ising-type transition from long-range order (LRO) to disordered state.
Since the $\theta$-field remains ordered, the system enters a vestigial charge-$4e$ SC phase. 
Upon further heating to $T/\rho\approx 0.549$, a second peak in $C_v$ appears. 
Here, $\chi_{\theta}$ diverges and the cumulant $3U_{\theta}-1$ rapidly vanishes, 
suggesting a 3D XY-type transition of the $\theta$-field.
Above this temperature, the system enters the normal metallic phase. 

(ii). For $\kappa/\rho=0.93$, the results are summarized in Fig.~\ref{Eg_chiral_ob}(a2,b2,...,g2).
A single simultaneous transition is observed at $T/\rho\approx 0.549$.
At this temperature, specific heat $C_v$ shows a pronounced peak. 
The susceptibilities $\chi_{\theta}$ and $\chi _{\phi}$ are divergent, 
while the cumulants $3U_{\theta}-1$ and $\frac{3}{2}U_{\phi}$, along with the phase stiffness $S$, drop sharply to zero.
This indicates a direct transition from the chiral SC state to the normal metal phase.

(iii). For $\kappa/\rho=1.2$, the resutls are summarized in Fig.~\ref{Eg_chiral_ob}(a3,b3,...,g3).
Clearly, the order of transitions is reversed compared to the $\kappa/\rho=0.6$ case.
At $T/\rho\approx 0.549$, the specific heat $C_v$ exhibits a peak. 
The $\theta$-field disorders first, evidenced by the susceptibility $\chi_{\theta}$ changing divergently, along with the drop in the cumulant $3U_{\theta}-1$ and the phase stiffness $S$.
The system thus loses SC and enters a chiral metal phase,
where $\phi$-field persists to be ordered.
Subsequently, at $T/\rho\approx 0.692$, the specific heat exhibits a new peak. 
The $\phi$-field also disorders, indicated by the divergence of the $\chi _{\phi}$, the vanishing of the Ising order parameter $I$ and $\frac{3}{2}U_{\phi}$.
Consequently, the $\phi$-field experiences an Ising phase transition
and the system enters the normal metallic phase.

\begin{table}[t]
	\centering
	\begin{tabular}{|c|c|c|c|c|c|c|}
		\hline\hline
		Phase & $G_\theta$ & $G_{\phi}$ \\
		\hline
		charge-$4e$ SC & ~~const~~ & ~~$e^{-r/\xi}$~~ \\
		\hline
		metal & ~~$e^{-r/\xi_1}$~~ & ~~$e^{-r/\xi_2}$~~ \\
		\hline
		chiral SC & ~~const & ~~const~~ \\
		\hline
		chiral metal & ~~$e^{-r/\xi}$~~ & ~~const~~ \\
		\hline\hline
	\end{tabular}
    \caption{The correlation functions $G_\theta$ and $G_{\phi}$ decay for all possible phases in Fig.~\ref{Eg_chiral}.  The abbreviations denote: chiral SC is chiral superconductivity; charge-$4e$ SC is charge $4e$ superconductivity; metal is normal metal; chiral metal is chiral metal.}\label{tab:1}
\end{table}

\begin{figure}[t]
\centering
\includegraphics[width=0.45\textwidth]{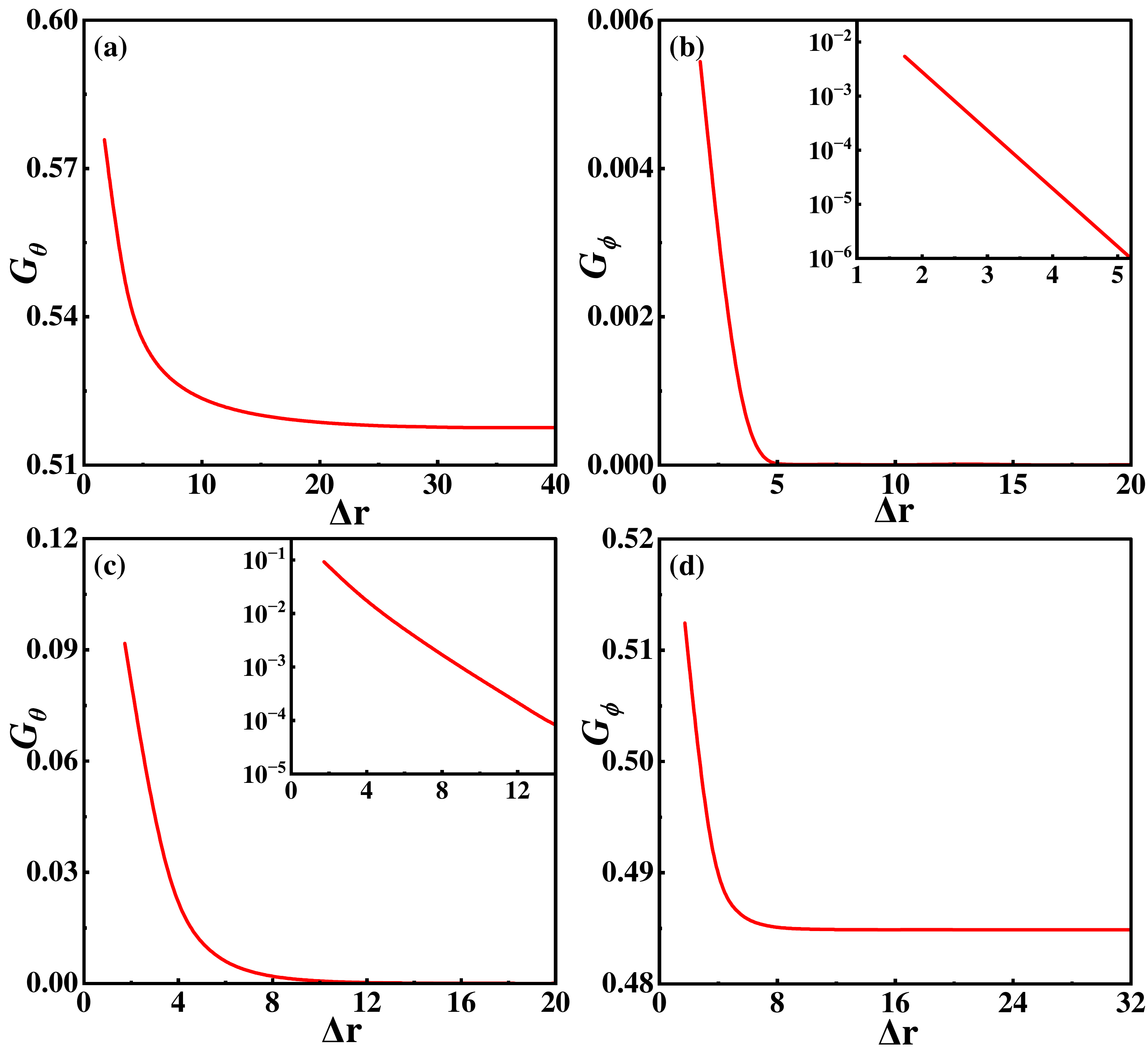}
\caption{(Color online) The correlation function $G_{\theta/\phi}$ for (a) and (b) for the point $\bm{A}$ ($\kappa=0.3\rho, T=0.4\rho$), for (c) and (d) for the point $\bm{D}$ ($\kappa=1.3\rho, T=0.6\rho$) labeled in Fig.~\ref{Eg_chiral}. Insets of (b)-(c) only the y-axis is logarithmic. The $G_{\theta/\phi}$ for $\bm{B}, \bm{C}$ are shown in the appendix.}
\label{Eg_chiral_AD}
\end{figure}

Tab.~\ref{tab:1} summarizes the asymptotic behavior of the correlation functions,
$G_{\theta}(\Delta \bm{r})$ and $G_{\phi}(\Delta \bm{r})$, 
for the distinct possible phases identified in Fig.~\ref{Eg_chiral}.
Here, we specifically examine the nature of the vestigial phases:
the charge-$4e$ SC and chiral metal phases. 
Fig.~\ref{Eg_chiral_AD}(a) and (b) illustrate the spatial dependence of the correlation function $G_{\theta}$ and $G_{\phi}$, as a function of the relative separation $\Delta r$ ($\equiv |\Delta \bm{r}|$), for the typical point $\bm{A}$ marked in Fig.~\ref{Eg_chiral}. 
It is evident that $G_{\theta}$ saturates to a finite value in the long-distance limit $\Delta r \to \infty$, indicating the presence of the LRO in the $\theta$-field.
In contrast, $G_{\phi}$ decays exponentially with $\Delta r$,
signifying that $\phi$-field is disordered.
This coexistence of the LRO $\theta$-field and disordered $\phi$-field suggests the realization of the charge-$4e$ SC. 
Conversely, for the representative point $\bm{D}$ in Fig.~\ref{Eg_chiral}, the correlation functions summarized in Fig.~\ref{Eg_chiral_AD}(c) and (d) exhibit reversed behavior.
As $\Delta r \to \infty$, $G_{\theta}$ displays exponential decay, while $G_{\phi}$ saturates to a finite value. 
This behavior reflects a disordered $\theta$-field coexisting with the LRO in the $\phi$ field, 
which is the defining characteristic of the chiral metal phase.

\subsection{$T_{2g}/T_{1u}$ representation}

We now turn our attention to the phase diagram for phase fluctuations problem of the chiral SC state within the $T_{2g}$ (and similarly $T_{1u}$) IRRP.
To facilitate MC simulations, we also discretize the continuum Hamiltonian (Eq.~(\ref{Hamiltonian_r_T2g})) onto a 3D cubic lattice.

Similar to the $E_g$ case, it is crucial to faithfully respect the topological structure of the phase space during the simulations. The system is fundamentally described by three independent SC phase components: $\theta_1$, $\theta_2$ and $\theta_3$.
While it is physically intuitive to decompose these into a global phase $(\theta=\frac{1}{3}\sum_a\theta_a)$ and relative phaes $(\phi_a=\theta_a-\theta)$, these composite variables exhibit complex, coupled periodicities and are subject to a strict zero-sum constraint $\sum_a\phi_a=0$.
Sampling these constrained variables directly is numerically complicated. 
In contrast, the original component phases $\theta_a$ are fundamental condensate phases that independently and strictly obey a simple modulo $2\pi$ periodicity.
Therefore, to unambiguously capture the underlying physics without introducing artificial gauge constraints, we directly select $\theta_1,\theta_2$ and $\theta_3$ as the fundamental degrees of freedom to be sampled in the interval $[0,2\pi)$.

To construct the lattice interactions using these independent variables, we relate the global phase mode to the sum of the components ($\sum_a\theta_a\leftrightarrow 3\theta$) and the relative phase modes to their pairwise differences ($\theta_a-\theta_b\leftrightarrow\phi_a-\phi_b$).
Applying the standard periodic discretization scheme as before, the lattice Hamiltonian is constructed as follows:
\begin{eqnarray}\label{Hamiltonian_d_T2g}
H &=& -\frac{\rho}{9}\sum_{\langle ij\rangle} \cos\big[ \sum_{a}\theta_a(\bm{r}_{i}) -\sum_a \theta_a(\bm{r}_{j}) \big]  \nonumber\\
&-& \frac{\kappa}{3} \sum_{\langle ij\rangle} \sum_{a<b} 
\cos\big[\theta_{a}(\bm{r}_{i})-\theta_{b}(\bm{r}_{i}) 
-\theta_{a}(\bm{r}_{j})+\theta_{b}(\bm{r}_{j}) \big] \nonumber\\
&+& A\sum_{i} \sum_{a<b} \cos\big[2\theta_a(\bm{r}_{i})-2\theta_b(\bm{r}_{i})\big],
\end{eqnarray}
with $a,b=1,2,3$.
Here, the summation of $\langle ij\rangle$ runs over all the nearest-neighbor bonds.
%The positive prefactors $\rho/9$ and $\kappa/3$ are specially chosen to ensure consistency with the continuum limit defined in Eq.~(\ref{Hamiltonian_r_T2g}).

To verify the consistency with the continuum limit, we substitute the relations $\sum_a \theta_a = 3\theta$ and $\theta_a - \theta_b = \phi_a - \phi_b$ back into Eq.~(\ref{Hamiltonian_d_T2g}). 
For the global phase term, expanding the cosine function in the slow-mode limit yields $-\frac{\rho}{9} \cos(3\Delta\theta_{ij}) \approx \text{const} + \frac{\rho}{2}(\Delta\theta_{ij})^2$, which simplifies to the required kinetic term. 
For the relative phase term, substituting the phase differences and expanding the cosine sum gives $\frac{\kappa}{6} \sum_{a<b} (\Delta\phi_{a,ij} - \Delta\phi_{b,ij})^2$. 
By utilizing the zero-sum constraint of the relative phases ($\sum_a \Delta\phi_{a,ij} = 0$), we apply the algebraic identity $\sum_{a<b} (x_a - x_b)^2 = 3\sum_a x_a^2$, which reduces the relative kinetic energy to $\frac{\kappa}{2} \sum_a (\Delta\phi_{a,ij})^2$. 
Meanwhile, the local anisotropy term straightforwardly evaluates to $A \sum_{a<b} \cos(2\phi_{a,i} - 2\phi_{b,i})$. 
Replacing the discrete bond differences with spatial gradients recovers the continuum Hamiltonian Eq.~(\ref{Hamiltonian_r_T2g}).

\begin{figure}[t]
\centering
\includegraphics[width=0.3\textwidth]{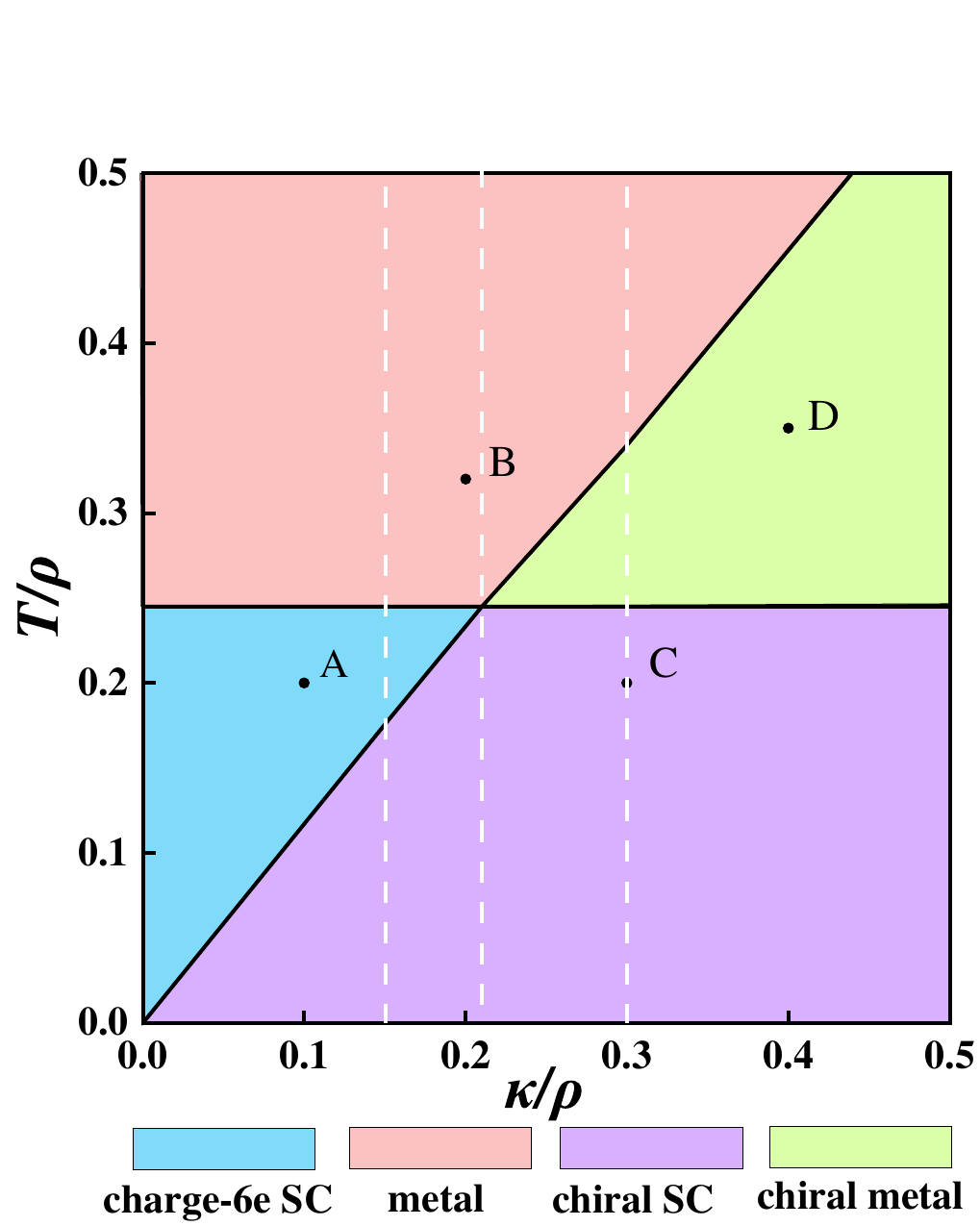}
\caption{(Color online) Phase diagram provided by the MC study for the chiral SC with the order parameters in the $T_{2g}/T_{1u}$ representations with $A=0.16\rho$ in Eq.~(\ref{Hamiltonian_d_T2g}). The white dashed lines mark $\kappa/\rho=0.15, 0.21$ and $0.3$, respectively.}\label{T2g_chiral}
\end{figure}

\begin{figure}[t]
	\centering
	\includegraphics[width=0.43\textwidth]{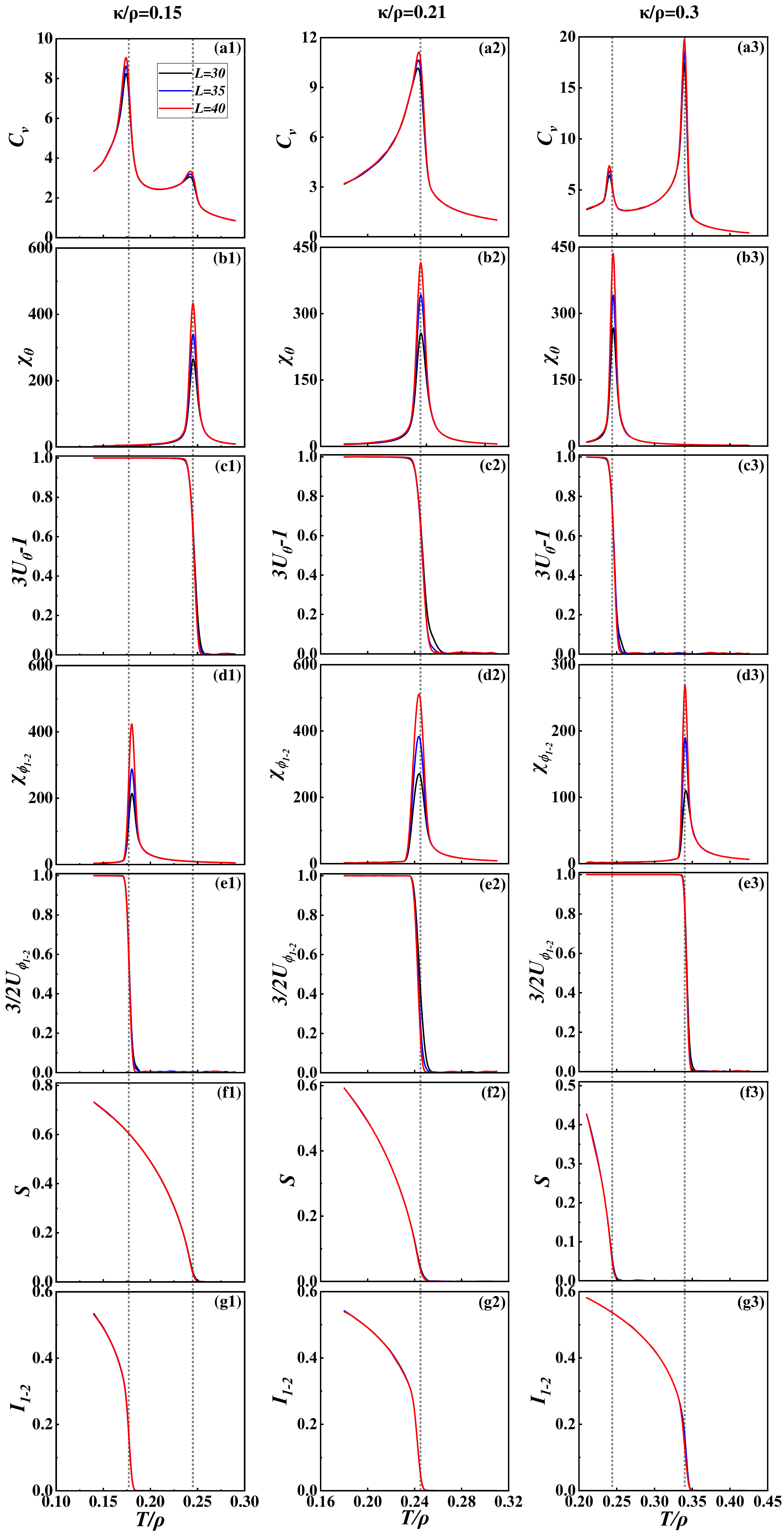}
	\caption{(Color online) The quantities as functions of temperature for $\kappa/\rho=0.15$ (a1,b1,...,g1), $\kappa/\rho=0.21$ (a2,b2,...,g2) and $\kappa/\rho=0.3$ (a3,b3,...,g3) (the $T_{2g}/T_{1u}$ representations). The scaling in all figures is $L=$ 30 (black line), 35 (blue line), and 40 (red line). (a1-a3) The specific heat $C_v$. (b1-b3) The susceptibility $\chi_{\theta}$ of $\theta$ field. (c1-c3) $3U_{\theta}-1$, where $U_{\theta}$ is the Binder cumulant of the $\theta$-field. (d1-d3) The susceptibility $\chi_{\phi_{1-2}}$ of $\phi_1-\phi_2$ field. (e1-e3) $\frac{3}{2}U_{\phi_{1-2}}$, where $U_{\phi_{1-2}}$ is the Binder cumulant of the $\phi_1-\phi_2$ field. (f1-f3) The phase stiffness $S$ of $\theta$ field. (g1-g3) The Ising order parameter $I_{1-2}$ of $\phi_1-\phi_2$ field. The grey dashed lines represent the phase transitions in (a1)-(g3).}\label{T2g_chiral_ob}
\end{figure}

We fix the anisotropic parameter $A=0.16\rho$ in the simulation (the case $A =0$ is also discussed in the appendix).
The phase diagram is depicted in Fig.~\ref{T2g_chiral}.
To elucidate the phase transition behavior, we calculate several key observables:
the specific heat $C_v$; the susceptibilities for the total phase $\chi_{\theta}$ and relative phases $\chi_{\phi_{1-2}}$; the Binder cumulants for the total phase $3U_{\theta}-1$ and relative phases $\frac{3}{2}U_{\phi_{1-2}}$; 
the phase stiffness $S$ associated with the $\theta$-field; 
and the Ising order parameter $I_{1-2}$ for the relative phase $\phi_1-\phi_2$.
Note that due to symmetry, the behavior of the three relative phase channels ($\phi_1-\phi_2$, $\phi_2-\phi_3$ and $\phi_3-\phi_1$) is consistent;
thus, we can focus on $\chi_{\phi_{1-2}}$, $\frac{3}{2}U_{\phi_{1-2}}$ and $I_{1-2}$ as representative quantities.
Fig.~\ref{T2g_chiral_ob} summarizes these observables for  lattice sizes $L=30, 35, 40$ at three representative coupling ratios: $\kappa/\rho = 0.15, 0.21, 0.3$, corresponding to the vertical cuts indicated in Fig.~\ref{T2g_chiral}. 
Comprehensive data for all observables are provided in the appendix.

Thermodynamic results in Fig.~\ref{T2g_chiral_ob} reveals three distinct transition scenarios depending on the stiffness ratio $\kappa/\rho$:
(i). For $\kappa/\rho=0.15$ (Fig.~\ref{T2g_chiral_ob}(a1,b1,...,g1)), the system undergoes two separate phase transitions upon heating. 
The first occurs at $T/\rho \approx 0.176$, 
where the divergent peak in $C_v$ and divergence of $\chi_{\phi_{1-2}}$ signal the disordering of the relative $\phi_1-\phi_2$ phase. 
The vanishing of the Ising order parameter $I_{1-2}$ and the cumulant $\frac{3}{2}U_{\phi_{1-2}}$ confirm that this transition belongs to the Ising universality class.
Since the $\theta$ field remains ordered in this regime, the system enters a vestigial charge-$6e$ SC phase.
A second transition follows at $T/\rho \approx 0.245$, marked by another peak in $C_v$.
Here, the susceptibility $\chi_{\theta}$ of $\theta$-field tends to divergence with vanishing $3U_{\theta}-1$, indicating the loss of the $\theta$-field coherence and the transition into the normal metal phase.

(ii). For $\kappa/\rho=0.21$ (Fig.~\ref{T2g_chiral_ob}(a2,b2,...,g2)), a single simultaneous transition is observed at $T/\rho \approx 0.245$. 
This is evidenced by a sharp peak in the specific heat $C_v$ and the concurrent vanishing of the phase stiffness $S$ and $3U_{\theta}-1$.
Simultaneously, $\chi_{\theta}$ and $\chi_{\phi_{1-2}}$ diverge.
This indicate both the global phase $\theta$ and relative phases $\phi_1-\phi_2$ become disordered at the same temperature, leading to a direct transition from chiral SC to the normal metal phase.

(iii). For $\kappa/\rho=0.3$ (Fig.~\ref{T2g_chiral_ob}(a3,b3,...,g3)), the order of transition is reversed compared to the $\kappa/\rho=0.15$ case. 
At $T/\rho \approx 0.245$, a peak in $C_v$ appears. 
The $\theta$-field disorders first at this point, 
indicated by the susceptibility $\chi_{\theta}$ becoming divergent with the vanishing phase stiffness $S$.
Thus, the system enters to a chiral metal phase, where relative phase order persists.
Subsequently, at $T/\rho \approx 0.34$, another $C_v$ peak occurs with divergent $\chi_{\phi_{1-2}}$, $\frac{3}{2}U_{\phi_{1-2}}$ and vanishing $I_{1-2}$.
The relative phases disorder, confirming the system undergoes an Ising transition of the $\phi_1-\phi_2$ field into the the normal metal phase.

\begin{table}[t]
	\centering
	\begin{tabular}{|c|c|c|c|c|c|c|}
		\hline\hline
		Phase & $G_\theta$ & $G_{\phi_{a-b}}$ \\
		\hline
		charge-$6e$ SC & ~~const~~ & ~~$e^{-r/\xi}$~~ \\
		\hline
		metal & ~~$e^{-r/\xi_1}$~~ & ~~$e^{-r/\xi_2}$~~ \\
		\hline
		chiral SC & ~~const & ~~const~~ \\
		\hline
		chiral metal & ~~$e^{-r/\xi}$~~ & ~~const~~ \\
		\hline\hline
	\end{tabular}	
    \caption{The correlation functions $G_\theta$ and $G_{\phi_{a-b}}$ decay for all possible phases in Fig.~\ref{T2g_chiral}. The abbreviations denote: charge-$6e$ SC is charge $6e$ superconductivity.}\label{tab:2}
\end{table}

\begin{figure}[t!]
\centering
\includegraphics[width=0.45\textwidth]{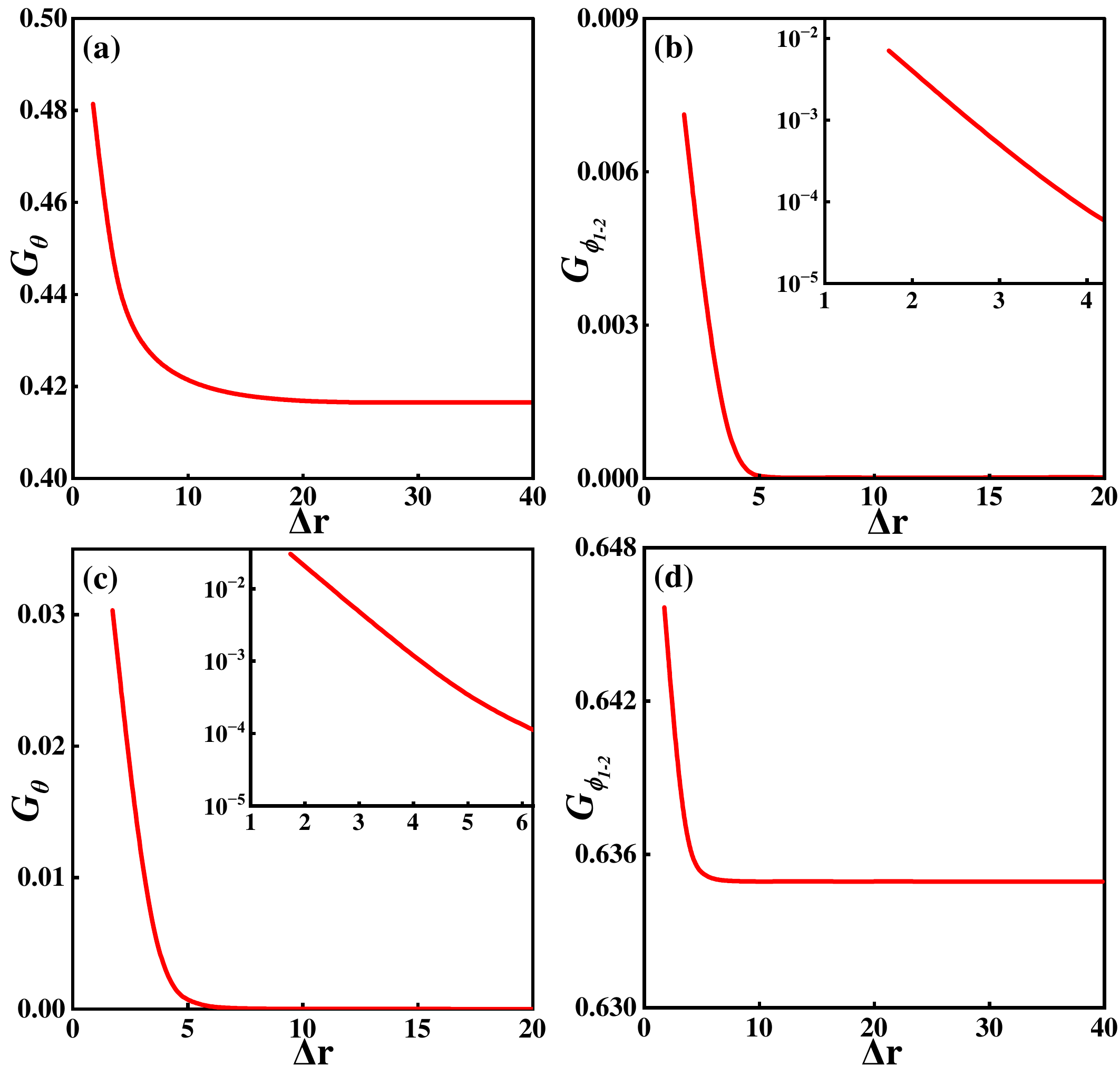}
\caption{(Color online) The correlation function $G_{\theta/\phi_{1-2}}$ for (a) and (b) for the point $\bm{A}$ ($\mu=0.1\rho, T=0.2\rho$), for (c) and (d) for the point $\bm{D}$ ($\mu=0.4\rho, T=0.35\rho$) labeled in Fig.~\ref{T2g_chiral}. Insets of (b)-(c) only the y-axis is logarithmic. The $G_{\theta/\phi_{a-b}}$ for $\bm{A}, \bm{B}, \bm{C}, \bm{D}$ are shown in the appendix.}
\label{T2g_chiral_AD}
\end{figure}

The nature of each phase is further corroborated by the asymptotic behavior of the spatial correlation functions, $G_{\theta}(\Delta \bm{r})$ and $G_{\phi_{a-b}}(\Delta \bm{r})$, which are summarized in Table~\ref{tab:2}. 
While only $G_{\phi_{1-2}}$ is shown in Fig.~\ref{T2g_chiral_AD} (with details for all $G_{\phi_{a-b}}$ provided in the appendix), these plots suffice to identify the phases. Specifically, a charge-$6e$ SC phase is identified at the representative point $\bm{A}$ in Fig.~\ref{T2g_chiral}. 
This is supported by panels (a) and (b) of Fig.~\ref{T2g_chiral_AD}.
Clearly, $G_{\theta}$ saturates to a constant value, confirming LRO in the $\theta$-field, and $G_{\phi_{1-2}}$ decays exponentially, indicating disorder in the $\phi_{a}-\phi_b$ field. 
On the other hand, the chiral metal phase is identified at point $\bm{D}$ in Fig.~\ref{T2g_chiral}, based on panels (c) and (d) of Fig.~\ref{T2g_chiral_AD}. 
There, $G_{\theta}$ decays exponentially, confirming disorder in the $\theta$-field, while $G_{\phi_{1-2}}$ saturates to a constant, confirming LRO in the $\phi_{a}-\phi_b$ field.

\section{Discussion and Conclusion}
\label{sec:CON}

In summary, we have investigated the emergence of vestigial phases in 3D multi-component superconductors under thermal melting, utilizing G-L symmetry argument and large-scale MC simulations on effective lattice models. 
The results reveal that the melting of the chiral SC state could exhibit distinct phase transition into the vestigial phases,
namely, charge-$4e/6e$ SC phase, chiral metal phase, and normal metallic state for the $E_g$, $T_{2g}$ and $T_{1u}$ representations. 
The transition between these phases is governed by the competition between the phase fluctuations in the total and relative phases channels,
capture by the corresponding phase stiffness $\rho$ and $\kappa$.
These findings provide a concrete theoretical framework and quantitative phase diagrams for searching such exotic states in 3D superconducting materials.

An important distinction emerges when comparing our 3D results with established 2D phenomena. 
In 2D systems, the melting of SC or chiral order is typically governed by the BKT mechanism, driven by the proliferation of topological vortex excitations. 
Specially, the proliferation of fractional vortice often induces a direct transition from chiral SC state to the normal state over a finite parameter range.
As a result, 2D phase diagrams generally feature two separated multicritical triple points, between which the two vestigial phases are disconnected by a direct transition line.

In contrast, our large-scale MC simulations on the 3D lattice reveal a qualitatively different scenario. 
Driven by conventional thermal fluctuations rather than vortex-mediated unbinding, the critical lines for global $U(1)$ and relative TRS breaking converge to a single intersection directly. 
We observe no finite window of direct transition that characterizes the 2D BKT-like behavior. 
Consequently, the 3D phase diagram features a single tetracritical point, where the chiral SC, vestigial SC, chiral metal, and normal state all coexist.

Furthermore, it is crucial to distinguish our findings from those in 3D systems governed by effective 2D point groups (such as tetragonal or hexagonal symmetries). 
While such lower-symmetry 3D systems may also exhibit tetracritical topologies due to the absence of BKT physics, they are usually restricted to two-component order parameters, thereby limiting their emergent vestigial states to charge-$4e$ SC.
By contrast, the 3D cubic symmetry ($O_h$ group) explored in our work might stabilize three-component $T_{2g}/T_{1u}$ IRRPs, 
giving rise to higher-order charge-$6e$ SC phase. 
These results highlight the roles of both dimensionality and spatial symmetry in determining the rich landscape of vestigial phases.

It is worth noting that our analysis of the tetracritical point relies on the assumption of weak coupling between the global and relative phase fluctuations. 
If significant inter-mode couplings, such as cross-gradient terms, are present, the mutual feedback between the two ordering channels could drive the disordering transition simultaneously.
This synergistic effect could merge the otherwise distinct phase boundaries into a single direct transition line bounded by multicritical triple points.
While this phenomenologically resembles the 2D phase diagram, it is driven by a fundamentally different 3D coupling mechanism rather than vortex unbinding.

Looking forward, while the present study focuses on the chiral SC ground states, our theoretical framework and lattice MC approach can be naturally extended to explore the thermal fluctuations in the nematic SC states. 
In a nematic superconductor, the system spontaneously breaks the discrete crystalline rotational symmetry rather than TRS. 
The partial melting of such an order driven by thermal fluctuations could stabilize distinct vestigial phases, most notably a nematic metal, a non-superconducting state that strongly breaks the lattice rotational symmetry. 
The resulting 3D phase diagram, governed by the competition between the global $U(1)$ phase stiffness and the discrete rotational symmetry fluctuations, is anticipated to exhibit its own unique multicritical topologies.

Looking forward, investigating how coupled fluctuations reshape the chiral phase diagram, as well as exploring the rich topologies of nematic vestigial phases, represent intriguing and promising directions for future research.

\section{Acknowledgments}

F.Y. is supported by the national natural science foundation of China under the Grant Nos.~12574141, 12234016 and 12074031.
Z.P. is supported by the National Natural Science Foundation of China under the Grants No.~12504219 and the startup funding from Xiamen University. 
C.L. is supported by the National Natural Science Foundation of China under the Grants No.~12304180.

%\onecolumngrid
\renewcommand{\thefigure}{S\arabic{figure}}
\renewcommand{\thetable}{S\arabic{table}}
\setcounter{figure}{0}
\setcounter{table}{0}

%\begin{widetext}

\appendix

\section{G-L analysis for $E_{g}$ chiral SC}

In the $E_g$ IRRP, the Ginzburg-Landau (G-L) free energy functional $F[\psi]$ is constructed based on the system's symmetries $G=O_h\times U(1)\times \mathcal{T}$,
where $O_h$ is the cubic point group, $U(1)$ denotes the global gauge symmetry, and $\mathcal{T}$ represents time reversal symmetry (TRS). 
Under the $O_h$ point group operations, the two-component order-parameter doublet $(\psi_1,\psi_2)$ transforms according to the IRRP $E_g$.
The specific transformation rules are summarized in the Tab.~\ref{tab:symmetry_rules}.

To systematically construct the invariant free energy terms, we analyze the tensor product of the order parameters. 
The product of the representation with itself decomposes as $E_g \otimes E_g = A_{1g} \oplus A_{2g} \oplus E_g$. 
This decomposition allows us to identify the independent symmetric bilinears formed by $\psi_a$ and $\psi_b$. 
The trivial identity channel ($A_{1g}$) corresponds to the total density $|\psi_1|^2+|\psi_2|^2$, while the antisymmetric channel ($A_{2g}$) generates the TRS breaking combinations $i(\psi_1^*\psi_2 -\psi_2^*\psi_1)$.
Constrained by these symmetries, the G-L free energy expansion up to the fourth order (Eq.~(\ref{eq:EgFreeEnergy})) is constructed from the invariants of these channels.

\begin{table}[t!]
\centering
\renewcommand{\arraystretch}{1.8} 
\begin{tabular}{l c}
\toprule
\makecell{\textbf{Symmetry} \\ \textbf{Operation}} & \makecell{\textbf{Order Parameters} \\ $(\psi_1, \psi_2) \to (\psi_1', \psi_2')$}  \\
\midrule
(1) $U(1)$ Gauge & $(e^{i\theta}\psi_{1}, e^{i\theta}\psi_{2})$  \\
\hline
(2) $C_4^1$ Rotation & \\
\quad (A) $C_{4z}^1$ ($z$-axis) & $(-\psi_{1}, \psi_{2})$  \\
\quad (B) $C_{4x}^1$ ($x$-axis) & $\left(\frac{1}{2}\psi_{1}-\frac{\sqrt{3}}{2}\psi_{2}, -\frac{\sqrt{3}}{2}\psi_{1}-\frac{1}{2}\psi_{2}\right)$  \\
\quad (C) $C_{4y}^1$ ($y$-axis) & $\left(\frac{1}{2}\psi_{1}+\frac{\sqrt{3}}{2}\psi_{2}, \frac{\sqrt{3}}{2}\psi_{1}-\frac{1}{2}\psi_{2}\right)$  \\
\hline
(3) $C_3^1$ (diag.) & $\left(-\frac{1}{2} \psi_{1}-\frac{\sqrt{3}}{2} \psi_{2}, \frac{\sqrt{3}}{2} \psi_{1}-\frac{1}{2} \psi_{2}\right)$  \\
\hline
(4) $C_2^1$ ($z$-axis) & $(\psi_{1}, \psi_{2})$  \\
\hline
(5) Inversion ($i$) & $(\psi_{1}, \psi_{2})$  \\
\hline
(6) Time-reversal & $(\psi_{1}^{\ast}, \psi_{2}^{\ast})$  \\
\bottomrule
\end{tabular}
\caption{Transformation rules for the $E_g$ order-parameter doublet $(\psi_1, \psi_2)$ under the symmetry operations of the group $O_{h} \times U(1) \times \text{TRS}$.}
\label{tab:symmetry_rules}
\end{table}

\subsection{$2$th G-L expansion for $E_{g}$ IRRP}
In this section, we construct the lowest-order gradient terms in the G-L free energy by contracting the IRRPs of the gradient tensor and the order parameter bilinears under the $O_h$ point group.
The spatial gradients (or momenta $\bm{k}$) transform under the $T_{1g}$ IRRP.
Therefore, the second-order gradient tensor $k_ik_j\sim \partial_i\partial_j$ decomposes under the $O_h$ group as,
\begin{align*}
T_{1u}\otimes T_{1u}
=A_{1g} \oplus E_g\oplus T_{2g}\oplus T_{1g}.
\end{align*}
The two relevant symmetric channels are the isotropic scalar channel $A_{1g}$ and the cubic anisotropic channel $E_g$.
Similarly, the tensor product of the two $E_g$ order parameters decomposes into:
\begin{align*}
E_g\otimes E_g
=A_{1g} \oplus A_{2g}\oplus E_g.
\end{align*}
To construct gradient terms that are fully invariant under the $O_h$ group, we should couple matching representations to realize the trivial identity representation $A_{1g}$.
This yields two independent allowed invariants corresponding to the $A_{1g}\otimes A_{1g}$ and $E_g\otimes E_g$ channels. 
Therefore, in momentum space, the general form of the second-order free energy requires two independent phenomenological stiffness coefficients, $K_1$ and $K_2$:
\begin{equation}
\begin{aligned}
&F_{0}^{(2)} = K_1 \bm{k}^2 \left( |\psi_1(\bm{k})|^2 + |\psi_2(\bm{k})|^2 \right) \\
& + K_2 \Bigg[ \frac{2k_z^2 - k_x^2 - k_y^2}{\sqrt{6}} \frac{|\psi_2(\bm{k})|^2 - |\psi_1(\bm{k})|^2}{\sqrt{2}} \\
&+ \frac{k_x^2 - k_y^2}{\sqrt{2}} \left( \psi_1^\ast(\bm{k})\psi_2(\bm{k}) + \psi_2^\ast(\bm{k})\psi_1(\bm{k}) \right) \Bigg]
\end{aligned}
\end{equation}
Here, the first term $K_1$ originates from the $A_{1g}$ channel and represents the standard, fully isotropic kinetic energy.
The second term $K_2$ originates from the $E_g$ channel and captures the inherent cubic anisotropy of the underlying lattice.

Transforming back to real space, the $A_{1g}$ contribution to the free energy takes the familiar isotropic form:
\begin{equation}
F_{A_{1g}}^{(2)} = K_1 \int d^3\bm{r} \left( |\nabla\psi_1|^2 + |\nabla\psi_2|^2 \right)
\end{equation}
To analyze the phase fluctuation regime, we substitute the chiral ansatz $\psi_{1,2}=\psi_0 e^{i(\theta\pm\phi)}$, assuming a uniform amplitude $\psi_0$.
For the isotropic $A_{1g}$ term, the gradient decoupling yields:
\begin{equation}
F_{A_{1g}}^{(2)} = 2K_1 \psi_0^2 \int d^3\bm{r} \left( |\nabla\theta|^2 + |\nabla\phi|^2 \right)
\end{equation}
If this were the only allowed term, the global phase $\theta$ and relative phase $\phi$ would exhibit stiffness degeneracy $\rho=\kappa$.

However, the chiral $d+id$ state explicitly breaks $C_4$ symmetry and lowers the macroscopic symmetry of the system from cubic to trigonal. 
Because the ground state does not retain macroscopic cubic isotropy, the anisotropic $E_g\otimes E_g$ channel governed by $K_2$ does not vanish. 
When the chiral phase ansatz is substituted into the $E_g$ gradient terms, it generates spatial cross-derivatives that couple the crystalline anisotropy directly to the phase modes. 
The $E_g$ channel lifts the degeneracy between the phase modes, yielding distinct effective stiffness coefficients for the global and relative phase fluctuations.

\subsection{$4$th G-L expansion for $E_{g}$ IRRP}
While the cubic anisotropy in the second-order gradient expansion already lifts the degeneracy between the global phase $\theta$ and relative phase $\phi$, higher-order gradient-field couplings provide an additional, density-dependent contribution to this stiffness splitting. 
To understand the origin of these corrections, it is instructive to analyze the symmetry properties of the fourth-order G-L terms via representation decomposition.

The general fourth-order invariant can be constructed by contracting a gauge-invariant order parameter bilinear (scaling as $\sim \psi^\ast \psi$) with a gauge-invariant gradient bilinear (scaling as $\sim \nabla\psi^\ast \cdot \nabla\psi$). 
Under the cubic point group $O_h$, the tensor product of the $E_g$ order parameter with its conjugate decomposes as $E_g \otimes E_g = A_{1g} \oplus A_{2g} \oplus E_g$. The corresponding gauge-invariant bilinears for the fields ($u$) and gradients ($g$) are:
isotropic $A_{1g}$ channels,
\begin{align}
u_0 = |\psi_1|^2 + |\psi_2|^2, \quad 
g_0 = |\nabla\psi_1|^2 + |\nabla\psi_2|^2
\end{align}
antisymmetric $A_{2g}$ channels,
\begin{equation}
\begin{aligned}
u_3 =& i(\psi_1^\ast\psi_2 - \psi_2^\ast\psi_1),  \\
g_3 =& i[(\nabla\psi_1^\ast)\cdot(\nabla\psi_2) - (\nabla\psi_2^\ast)\cdot(\nabla\psi_1)] 
\end{aligned}
\end{equation}
and anisotropic $E_g$ channels,
\begin{equation}
\begin{aligned}
&u_1 = |\psi_1|^2 - |\psi_2|^2, \quad g_1 = |\nabla\psi_1|^2 - |\nabla\psi_2|^2 \\
&u_2 = \psi_1^\ast\psi_2 + \psi_2^\ast\psi_1,  \\
&g_2 = (\nabla\psi_1^\ast)\cdot(\nabla\psi_2) + (\nabla\psi_2^\ast)\cdot(\nabla\psi_1)
\end{aligned}
\end{equation}
To form fully symmetric scalar invariants in the free energy, these bilinears must couple within their respective symmetry channels. 
The coupling in the trivial $A_{1g}$ channel ($u_0 g_0$) simply yields an isotropic, density-dependent renormalization of the standard kinetic energy, which maintains $\rho = \kappa$.

However, anisotropic stiffness corrections naturally emerge from the $A_{2g} \otimes A_{2g}$ and $E_g \otimes E_g$ channels. 
The symmetry-allowed fourth-order invariants from these channels are parameterized by phenomenological constants $b_1$ and $b_2$:
\begin{equation}
F_{0}^{(4)} \supset \int d^{3}\bm{r} \Big[ b_1 (u_3 g_3) + b_2 (u_1 g_1 + u_2 g_2) \Big].
\end{equation}
We can evaluate these invariants by substituting the chiral $d+id$ ansatz $\psi_{1,2} = \psi_0 e^{i(\theta \pm \phi)}$. 
Under this ansatz, the amplitude is uniform ($|\psi_1|^2 = |\psi_2|^2 = \psi_0^2$), causing $u_1 = 0$. 
The remaining field bilinears evaluate to $u_2 = 2\psi_0^2 \cos(2\phi)$ and $u_3 = 2\psi_0^2 \sin(2\phi)$. 
The gradient bilinears decouple into the phase fields as:
\begin{equation}
\begin{aligned}
&g_2 = 2\psi_0^2 (|\nabla\theta|^2 - |\nabla\phi|^2) \cos(2\phi),    \\ 
&g_3 = 2\psi_0^2 (|\nabla\theta|^2 - |\nabla\phi|^2) \sin(2\phi).
\end{aligned}
\end{equation}
Substituting these into the free energy invariants, we uncover a simplification. 
Assuming $b_1 = b_2 = b'$ for a locally approximately isotropic interaction limit, the trigonometric factors cleanly cancel ($\sin^2(2\phi) + \cos^2(2\phi) = 1$), yielding a pure stiffness-splitting term:
\begin{equation}
\begin{aligned}
F_{0}^{(4)} &\supset b' \int d^3\bm{r} \Big( u_3 g_3 + u_2 g_2 \Big) \\
&= 4b'\psi_{0}^{4} \int d^{3}\bm{r} \Big( |\nabla\theta|^{2} - |\nabla\phi|^{2} \Big).
\end{aligned}
\end{equation}
This demonstrates how the contraction of higher-order gradient and field bilinears contributes a density-dependent splitting to the phase modes. 
Including corrections from both the second-order crystalline anisotropy and these fourth-order couplings, the effective phase fluctuation Hamiltonian takes the form:
\begin{equation}
H_{0}=\int d^{3}\bm{r}\left( \frac{\rho}{2}|\nabla\theta|^{2}+\frac{\kappa}{2}|\nabla\phi|^{2} \right),
\end{equation}
where it is guaranteed by symmetry that $\rho \neq \kappa$.

\section{G-L analysis for $T_{2g}$ chiral SC}

\begin{table}[t!]
\centering
\renewcommand{\arraystretch}{1.5} 
\begin{tabular}{l c}
\toprule
\makecell{\textbf{Symmetry} \\ \textbf{Operation}} & \makecell{\textbf{Order Parameters} \\ $(\psi_1, \psi_2, \psi_3)$}  \\
\midrule
(1) $U(1)$ Gauge & $(e^{i\theta}\psi_1, e^{i\theta}\psi_2, e^{i\theta}\psi_3)$  \\
\hline
(2) $C_4^1$ Rotation & \\
\quad (A) $C_{4x}^1$ ($x$-axis) & $(-\psi_3, -\psi_2, \psi_1)$  \\
\quad (B) $C_{4y}^1$ ($y$-axis) & $(\psi_2, -\psi_1, -\psi_3)$  \\
\quad (C) $C_{4z}^1$ ($z$-axis) & $(-\psi_1, \psi_3, -\psi_2)$  \\
\hline
(3) $C_3^1$ Rotation & $(\psi_3, \psi_1, \psi_2)$  \\
\hline
(4) $C_2^1$ Rotation ($x$-axis) & $(-\psi_1, \psi_2, -\psi_3)$  \\
\hline
(5) Inversion ($i$) & $(\psi_1, \psi_2, \psi_3)$  \\
\hline
(6) Time reversal & $(\psi_1^*, \psi_2^*, \psi_3^*)$  \\
\bottomrule
\end{tabular}
\caption{Transformation rules for the order parameters $\psi_a$ (transforming as $\{d_{xy}, d_{yz}, d_{xz}\}$) under the symmetry operations of the $O_h$ point group, $U(1)$ gauge, and time reversal.}
\label{tab:symmetry}
\end{table}

We investigate a superconducting state characterized by a three-component order parameter $\psi_a$ ($a=1,2,3$) for $T_{2g}$ IRRP.
The specific transformation rules for these order parameters $\psi_a$ under the symmetry operations are summarized in Tab.~\ref{tab:symmetry}.
The free energy of the system must be invariant under the global $U(1)$ gauge symmetry, TRS, and the $O_h$ point group operations. 
Constrained by the symmetry group operations derived above, we can construct the G-L free energy functional up to the fourth order in the order parameters $\psi_a$, as shown in Eq.~(\ref{eq:TgFreeEnergy}).

\subsection{$2$th G-L expansion for $T_{2g}$ IRRP}

In this section, we construct the lowest-order gradient terms in the G-L free energy by contracting the IRRPs of the gradient tensor and the order parameter bilinears under the $O_h$ point group.
The spatial gradients transform under the $T_{1u}$ IRRP, meaning the second-order gradient tensor $k_\mu k_\nu \sim -\partial_\mu \partial_\nu$ decomposes as:
\begin{equation}
T_{1u} \otimes T_{1u} = A_{1g} \oplus E_{g} \oplus T_{2g} \oplus T_{1g}.
\end{equation}
Simultaneously, the product of the three-component $T_{2g}$ order parameter with its conjugate decomposes into:
\begin{equation}
T_{2g} \otimes T_{2g} = A_{1g} \oplus E_{g} \oplus T_{2g} \oplus T_{1g}.
\end{equation}
To construct a fully symmetric scalar invariant ($A_{1g}$ channel), we contract the matching symmetry channels. 
Because there are three symmetric channels shared between the gradients and the fields ($A_{1g}$, $E_g$, and $T_{2g}$), the most general G-L free energy at the quadratic level requires three independent stiffness coefficients ($K_1$, $K_2$, and $K_3$). 
In real space, this general $O_h$-invariant quadratic free energy is expressed as:
\begin{equation}
\begin{aligned}
&F_{0}^{(2)} = \int d^3\bm{r} \Big[ K_1 \sum_{a=1}^3 |\nabla \psi_a|^2 + K_2 \Big( |\partial_z \psi_1|^2 + |\partial_x \psi_2|^2  \\
& + |\partial_y \psi_3|^2 \Big)
+ K_3 \Big( (\partial_x \psi_1^\ast)(\partial_y \psi_3) + \text{h.c.} + \dots \Big) \Big],
\end{aligned}
\end{equation}
where $\psi_1, \psi_2, \psi_3$ correspond to the $d_{xy}, d_{yz}, d_{xz}$ basis components, respectively. 
The $K_1$ term represents the fully isotropic $A_{1g}$ kinetic energy, the $K_2$ term captures the anisotropic diagonal stiffness dictated by the specific orbital orientation ($E_g$ channel), and the $K_3$ term governs the symmetry-allowed off-diagonal mixing ($T_{2g}$ channel).

We focus on the physically dominant isotropic channel. 
Setting $K_2 = K_3 = 0$ and defining $K_1 = A$, the free energy simplifies to:
\begin{equation}
F_{0}^{(2)} \approx A \int d^3\bm{r} \sum_{a=1}^3 |\nabla\psi_a(\bm{r})|^2.
\end{equation}
To analyze the phase modes in this isotropic limit, we parameterize the order parameter with a uniform amplitude as $\psi_a = \psi_0 e^{i(\theta+\phi_a)}$, 
where $\theta$ is the global $U(1)$ phase and $\phi_a$ represents the relative phase fluctuations subject to the constraint $\sum_a \phi_a = 0$. 
Substituting this ansatz into the simplified free energy yields:
\begin{equation}
\begin{aligned}
F_{0}^{(2)} &= A\psi_{0}^{2} \int d^{3}\bm{r} \sum_{a=1}^{3} |\nabla (\theta+\phi_a)|^2 \\
&= A\psi_{0}^{2} \int d^{3}\bm{r} \Big( 3|\nabla\theta|^{2} + \sum_{a=1}^{3}|\nabla\phi_{a}|^{2} \Big),
\end{aligned}
\end{equation}
where the cross-terms $2\nabla\theta \cdot \nabla\phi_a$ naturally vanish upon summation precisely due to the constraint $\sum_a \nabla\phi_a = \nabla(\sum_a \phi_a) = 0$.

Comparing this to the standard effective phase fluctuation Hamiltonian,
\begin{equation}
H_{0} = \int d^{3}\bm{r} \left( \frac{\rho}{2}|\nabla\theta|^{2} + \frac{\kappa}{2} \sum_{a=1}^{3}|\nabla\phi_{a}|^{2} \right),
\end{equation}
we obtain $\rho = 6A\psi_0^2$ and $\kappa = 2A\psi_0^2$. Thus, in the purely isotropic limit, the stiffness ratio for the global and relative phase modes is analytically exactly $\rho/\kappa = 3$.

\subsection{$4$th G-L expansion for $T_{2g}$ IRRP}

While the second-order gradient isotropic expansion establishes a fixed ratio between the stiffness coefficients of the global phase $\theta$ and the relative phases $\phi_a$ (specifically, $\rho/\kappa=3$), 
this constraint is lifted by including anisotropic and higher-order contributions. 
Here, we consider the expansion of the free energy $F$ to the fourth order.

Analogous to the $E_g$ case, the non-trivial stiffness splitting could arise from inter-component gradient couplings. 
Instead of coupling gauge-invariant densities ($\sim \psi^\ast\psi$), we can construct the stiffness-splitting invariant by contracting the local pairing field bilinear $\Delta_{ab} = \psi_a \psi_b$ and its conjugate gradient bilinear $D_{ab}^\ast = (\nabla\psi_a^\ast)\cdot(\nabla\psi_b^\ast)$, where we assume an isotropic spatial contraction for the gradient operators ($A_{1g}$ in spatial indices). 
Under the cubic point group $O_h$, the tensor product of the $T_{2g}$ order parameter with itself decomposes as:
\begin{equation}
T_{2g} \otimes T_{2g} = A_{1g} \oplus E_{g} \oplus T_{1g} \oplus T_{2g}.
\end{equation}
The symmetric off-diagonal components of these bilinears specifically span the $T_{2g}$ irreducible subspace. 
That is, the triplets $(\psi_2\psi_3, \psi_3\psi_1, \psi_1\psi_2)$ and $(D_{23}^\ast, D_{31}^\ast, D_{12}^\ast)$ both transform as $T_{2g}$. 
To form a fully symmetric scalar invariant ($A_{1g}$) under both $O_h$ and global $U(1)$ gauge transformations, we can contract the $T_{2g}$ channel of the fields with the corresponding $T_{2g}$ channel of the gradients.

Guided by this representation analysis, the following fourth-order invariant form in momentum space is symmetry-allowed:
\begin{equation*}
\begin{aligned}
\sim \big( b'(\bm{k}_{1}\cdot\bm{k}_{2}+\bm{k}_{3}\cdot \bm{k}_{4}) \big)
\Big[&\psi_{1}^{\ast}(\bm{k}_{1})\psi_{2}^{\ast}(\bm{k}_{2})\psi_{1}(\bm{k}_{3})\psi_{2}(\bm{k}_{4}) \\
+&\psi_{1}^{\ast}(\bm{k}_{1})\psi_{3}^{\ast}(\bm{k}_{2})\psi_{1}(\bm{k}_{3})\psi_{3}(\bm{k}_{4})    \\
+&\psi_{2}^{\ast}(\bm{k}_{1})\psi_{3}^{\ast}(\bm{k}_{2})\psi_{2}(\bm{k}_{3})\psi_{3}(\bm{k}_{4}) \Big].
\end{aligned}
\end{equation*}
Transforming to real space and substituting the chiral ansatz, 
the gradient term becomes
\begin{equation}
\begin{aligned}
\sim &-b'\int d^{3}\bm{r} \sum_{a<b} \Big[ (\nabla\psi_{a}^{\ast})\cdot(\nabla\psi_{b}^{\ast}) \psi_{a}\psi_{b} +\text{h.c.} \Big]  \\
=&6b'\psi_{0}^{4} \int d^{3}\bm{r}|\nabla\theta|^{2}
-b'\psi_{0}^{4}\int d^{3}\bm{r}\sum_{a=1}^3 |\nabla\phi_a|^{2}.
\end{aligned}
\end{equation}
Unlike the second-order term which scales as $3|\nabla\theta|^2+\sum|\nabla\phi_a|^2$, this contribution modifies the stiffnesses with a different ratio $6:(-1)$,
effectively breaking the fixed proportionality $\rho/\kappa=3$.
The resulting effective Hamiltonian is:
\begin{equation}
H_{0}=\int d^{3}\bm{r}\left( \frac{\rho}{2}|\nabla\theta|^{2}
+\frac{\kappa}{2} \sum_a |\nabla\phi_a|^{2}\right),
\end{equation}
where generally $\rho/\kappa\neq 3$.

\section{G-L analysis for $T_{1u}$ chiral SC}

We define the superconducting gap functions using the IRRP $T_{1u}$ triplet basis $(p_x,p_y,p_z)$, which transforms as a vector under the cubic point group $O_h$. 
The general order parameter is expressed as the linear combination 
\begin{align}
\Delta(\bm{k}) =\psi_1 \Delta_1(\bm{k}) +\psi_2 \Delta_2(\bm{k}) +\psi_3 \Delta_3(\bm{k}).
\end{align}
The complex coefficients $\psi_a$ act as the order parameters, transforming according to the $T_{1u}$ IRRP (odd parity). 
The specific transformation rules for both the order parameters and the momentum vector $\bm{k}=(k_x,k_y,k_z)$ under the relevant symmetry operations are summarized in Table \ref{tab:p_wave_symmetry}.

\begin{table}[t!]
\centering
\renewcommand{\arraystretch}{1.5} % 
\begin{tabular}{l c}
\toprule
\makecell{\textbf{Symmetry} \\ \textbf{Operation}} & \makecell{\textbf{Order Parameters} \\ $(\psi_1, \psi_2, \psi_3)$}  \\
\midrule
(1) $U(1)$ Gauge & $(e^{i\theta}\psi_1, e^{i\theta}\psi_2, e^{i\theta}\psi_3)$  \\
\hline
(2) $C_4^1$ Rotation & \\
\quad (A) $C_{4x}^1$ ($x$-axis) & $(\psi_1, \psi_3, -\psi_2)$  \\
\quad (B) $C_{4y}^1$ ($y$-axis) & $(-\psi_3, \psi_2, \psi_1)$  \\
\quad (C) $C_{4z}^1$ ($z$-axis) & $(-\psi_2, \psi_1, \psi_3)$  \\
\hline
(3) $C_3^1$ Rotation & $(\psi_3, \psi_1, \psi_2)$  \\
\hline
(4) $C_2^1$ Rotation ($x$-axis) & $(\psi_1, -\psi_2, -\psi_3)$  \\
\hline
(5) Inversion ($i$) & $(-\psi_1, -\psi_2, -\psi_3)$  \\
\hline
(6) TRS & $(\psi_1^*, \psi_2^*, \psi_3^*)$  \\
\bottomrule
\end{tabular}
\caption{Transformation rules for the vector order parameters $(\psi_1, \psi_2, \psi_3)$ (representing $\{p_x, p_y, p_z\}$ symmetry) under the symmetry group $G = O_h \times U(1) \times \text{TRS}$.}
\label{tab:p_wave_symmetry}
\end{table}

Finally, it is instructive to note that the G-L analysis for the $T_{1u}$ IRRP is formally analogous to that of the $T_{2g}$ IRRP discussed previously. 
Although the basis functions differ in parity ($p$-wave vs. $d$-wave), both are three-dimensional IRRPs of the $O_h$ group and share identical tensor product decomposition rules (i.e., the structure of their invariant scalars is the same), $T_{1u}\otimes T_{1u}=A_{1g} \oplus E_g \oplus T_{1g} \oplus T_{2g}$ and $T_{2g}\otimes T_{2g}=A_{1g} \oplus E_g \oplus T_{1g} \oplus T_{2g}$.
Consequently, the resulting form of the free energy expansion up to the fourth order for the $T_{1u}$ IRRP is identical to that of the $T_{2g}$ case.

\section{Detailed Results for MC study of $E_{g}$ chiral SC}

\begin{figure}[t!]
\centering
\includegraphics[width=0.9\linewidth]{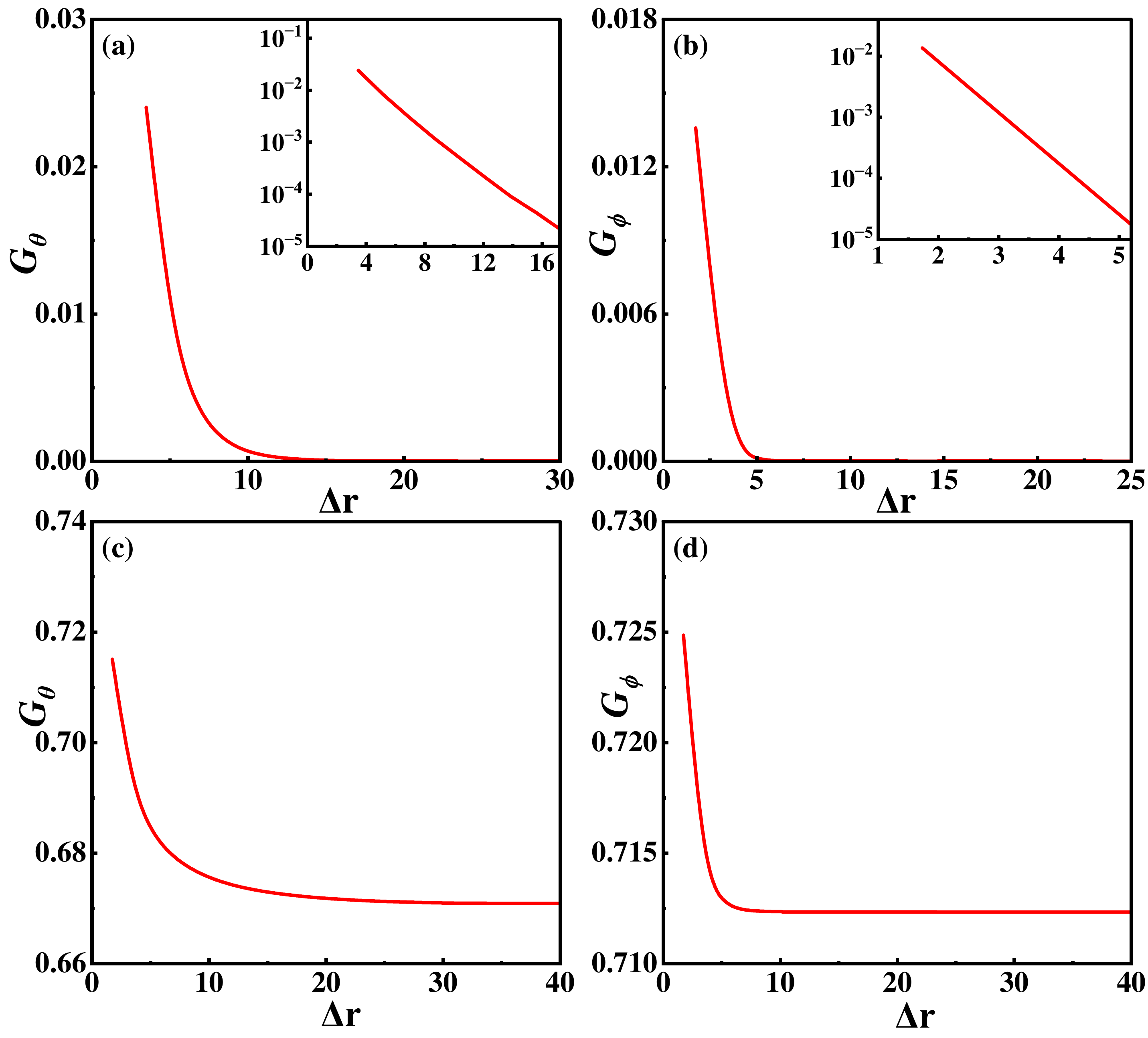}
\caption{(Color online) The correlation function $G_{\theta/\phi}$ (the $E_{g}$ representation) for (a) and (b) for the point $\bm{B}$ ($\kappa=0.6\rho, T=0.6\rho$), for (c) and (d) for the point $\bm{C}$ ($\kappa=0.9\rho, T=0.3\rho$) labeled in Fig.~\ref{Eg_chiral}. Insets of (a)-(b) only the y-axis is logarithmic.}
\label{Eg_chiral_BC}
\end{figure}

For the $E_g$ chiral SC system in Eq.~(\ref{Hamiltonian_d_Eg}), we characterize the phase transitions using the following thermodynamic observables and correlation functions. 
The specific heat per site is calculated from the energy fluctuations:
\begin{align}
C_v=\frac{1}{NT^2} \Big( \langle H^2 \rangle - \langle H \rangle^2 \Big),
\end{align}
where $N$ denotes the total number of lattice sites, and $H$ is the total Hamiltonian.
To probe the superfluid response, we compute the superfluid stiffness (or helicity modulus) associated with the total phase $\theta$ along the $x$-direction~\cite{zeng2024high}:
\begin{align}
S_{x}=\frac{1}{N} \Big( \langle H_{x}\rangle -\beta \langle I_{x}^2\rangle \Big),
\end{align}
where $\beta=1/k_BT$ and 
\begin{equation}
\begin{aligned}
H_x &= \rho\sum_{\langle ij\rangle_x}\cos[\theta_{1}(\bm{r}_i)+\theta_{2}(\bm{r}_i)-\theta_{1}(\bm{r}_j)-\theta_{2}(\bm{r}_j)], \\
I_x &= \rho\sum_{\langle ij\rangle_x}\sin[\theta_{1}(\bm{r}_i)+\theta_{2}(\bm{r}_i)-\theta_{1}(\bm{r}_j)-\theta_{2}(\bm{r}_j)].
\end{aligned}
\end{equation}
To characterize the relative-phase ordering, we calculate the Ising-type order parameter:
\begin{equation}
I\equiv\frac{1}{N^2}\sum_{ij}\left\langle\sin[\theta_{1}(\bm{r}_{i})-\theta_{2}(\bm{r}_{i})]\cdot \sin[\theta_{1}(\bm{r}_{j})-\theta_{2}(\bm{r}_{j})]\right\rangle.
\end{equation}

We further analyze the phase transitions by computing the susceptibility $\chi$~\cite{janke1990}: 
\begin{eqnarray}
\chi_{\theta}&=\dfrac{N(\left\langle m^2\right\rangle-\left\langle |m|\right\rangle^2)}{K_BT}, \\
\chi_{\phi}&=\dfrac{N(\left\langle m^2\right\rangle-\left\langle m\right\rangle^2)}{K_BT},
\end{eqnarray}
and the Binder cumulant $U$ of $\theta$ and $\phi$ fields:
\begin{eqnarray}
U=1-\dfrac{\left\langle m^4\right\rangle}{3\left\langle m^2\right\rangle^2}.
\end{eqnarray}
Here, the order parameters for the $\theta$ and $\phi$ fields are defined as: 
\begin{equation}
\begin{aligned}
m_{\theta}=&\frac{1}{N}\sum_ie^{i(\theta_1(\bm{r}_{i})+\theta_2(\bm{r}_{i}))},   \\
m_{\phi}=&\frac{1}{N}\sum_i \text{Im}(e^{i(\theta_1(\bm{r}_{i})-\theta_2(\bm{r}_{i}))}).
\end{aligned}
\end{equation}
To account for different universality classes, we use the normalized Binder ratios:
$U_{\theta}^*=3U_{\theta}-1$ for the XY transtion and $U_{\phi}^*=\frac{3}{2}U_{\phi}$ for the Ising universality class.

Finally, the spatial decay of correlations is characterized by the two-point correlation functions:
\begin{equation*}
G_{\theta/\phi}(\Delta \bm{r})\equiv\frac{1}{N}\sum_{\bm{r}}\left\langle e^{i[\theta_1(\bm{r}) \pm \theta_2(\bm{r})-\theta_1(\bm{r}+\Delta \bm{r}) \mp\theta_2(\bm{r}+\Delta \bm{r})]}\right\rangle,  
\end{equation*}
where $\Delta\bm{r}$ is the relative distance.

For the typical points marked in Fig.~\ref{Eg_chiral}, the spatial correlation functions exhibit distinct behaviors indicative of different phases. At point $\bm{B}$, both $G_{\theta}$ and $G_{\phi}$ decay exponentially with $\Delta r$ (Fig.~\ref{Eg_chiral_BC}(a)-(b)), which is consistent with the metal phase. In contrast, at point $\bm{C}$, these functions saturate to a finite value as $\Delta r \to \infty$ (Fig.~\ref{Eg_chiral_BC}(c)-(d)), consistent with the chiral SC.

\section{Detailed Results for MC study of $T_{2g}/T_{1u}$ chiral SC}

\begin{figure}[t!]
	\centering
	\includegraphics[width=0.85\linewidth]{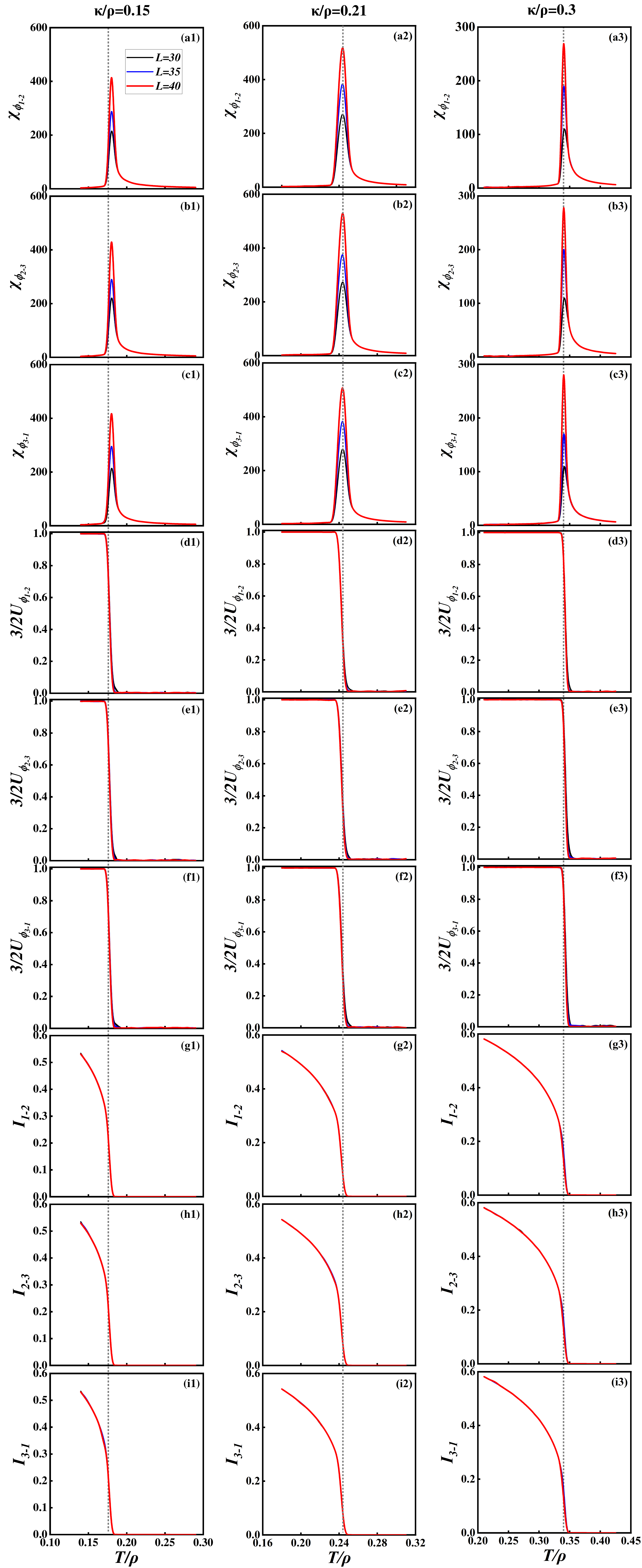}
	\caption{(Color online) The quantities as functions of temperature for $\kappa/\rho=0.15$ (a1,b1,..,i1), $\kappa/\rho=0.21$ (a2,b2,...,i2) and $\kappa/\rho=0.3$ (a3,b3,...,i3) (the $T_{2g}/T_{1u}$ representations with $A=0.16\rho$). The scaling in all figures is $L=$ 30 (black line), 35 (blue line), and 40 (red line). (a1-a3), (b1-b3) and (c1-c3) The susceptibilities $\chi_{\phi_{a-b}}$ of $\phi_a-\phi_b$ field, respectively. (d1-d3), (e1-e3) and (f1-f3) $\frac{3}{2}U_{\phi_{a-b}}$, where $U_{\phi_{a-b}}$ is the Binder cumulant of the $\phi_a-\phi_b$ field, respectively. (g1-g3), (h1-h3) and (i1-i3) The Ising order parameters $I_{a-b}$ of $\phi_a-\phi_b$ field, respectively. The grey dashed lines represent the phase transitions in (a1)-(i3).}\label{T2g_chiral_ob_2}
\end{figure}

For the chiral SC within three-dimensional IRRPs $T_{2g}/T_{1u}$,
we consider the discretized Hamiltonian in Eq.~(\ref{Hamiltonian_d_T2g}).
The associated phase stiffness characterizing the long-range order of the $\theta$ field and superconducting phase is
\begin{eqnarray}
S=\frac{1}{N}(\langle H_x\rangle -\beta\langle I_x^2\rangle),
\end{eqnarray}
with
\begin{equation}
\begin{aligned}
H_x &= \rho\sum_{\langle ij\rangle_x} \cos\big(\sum_a\theta_a(\bm{r}_i)
-\sum_a\theta_a(\bm{r}_j)\big),     \\
I_x &= \rho\sum_{\langle ij\rangle_x} \sin\big(\sum_a\theta_a(\bm{r}_i)
-\sum_a\theta_a(\bm{r}_j)\big),
\end{aligned}
\end{equation}
where $\beta=1/k_BT$.
The Ising order parameters are used to quantify the relative-phase ordering,
\begin{equation}
I_{a-b} =\frac{1}{N^2}\sum_{ij}\left\langle\sin[\theta_{a}(\bm{r}_{i})-\theta_{b}(\bm{r}_{i})] \sin[\theta_{a}(\bm{r}_{j})-\theta_{b}(\bm{r}_{j})]\right\rangle. 
\end{equation}

The susceptibilities $\chi$ are  
\begin{eqnarray}
\chi_{\theta}&=\dfrac{N(\left\langle m^2\right\rangle-\left\langle |m|\right\rangle^2)}{K_BT},\\
\chi_{\phi_{a}-\phi_{b}}&=\dfrac{N(\left\langle m^2\right\rangle-\left\langle m\right\rangle^2)}{K_BT},
\end{eqnarray}
and the Binder cumulant $U$ of $\theta$ and $\phi_{a}-\phi_{b}$ fields are given as
\begin{eqnarray}
U=1-\dfrac{\left\langle m^4\right\rangle}{3\left\langle m^2\right\rangle^2},
\end{eqnarray}
where $m_{\theta}=\frac{1}{N}\sum_ie^{i(\theta_1(\bm{r}_i)+\theta_2(\bm{r}_i)+\theta_3(\bm{r}_i))}$ for the $\theta$-field or $m_{\phi_{a-b}}=\frac{1}{N}\sum_i \text{Im} (e^{i(\theta_a(\bm{r}_{i})-\theta_b(\bm{r}_{i}))})$ for the $\phi_{a}-\phi_{b}$ field. The normalized Binder ratio takes distinct values for different universality classes: it is $3U_{\theta}-1$ for the XY-type transition in the $\theta$ field, and $\frac{3}{2}U_{\phi_{a-b}}$ for the Ising-type transition in the $\phi_a-\phi_b$ field.

The $\theta$ and $\phi_{a}-\phi_{b}$ fields correlation functions are defined as
\begin{equation}
\begin{aligned}
G_{\theta}(\Delta \bm{r})&=\frac{1}{N}\sum_{\bm{r}}\left\langle e^{i(\sum_a\theta_a(\bm{r})- \sum_a\theta_a(\bm{r}+\Delta \bm{r}))}\right\rangle,  \nonumber\\
G_{\phi_{a-b}}(\Delta \bm{r})&=\frac{1}{N}\sum_{\bm{r}}\left\langle e^{i(\theta_{a}(\bm{r})-\theta_{b}(\bm{r})-\theta_{a}(\bm{r}+\Delta \bm{r})+\theta_{b}(\bm{r}+\Delta \bm{r}))}\right\rangle.  
\end{aligned}
\end{equation}

\begin{figure}[t!]
	\centering
	\includegraphics[width=0.4\textwidth]{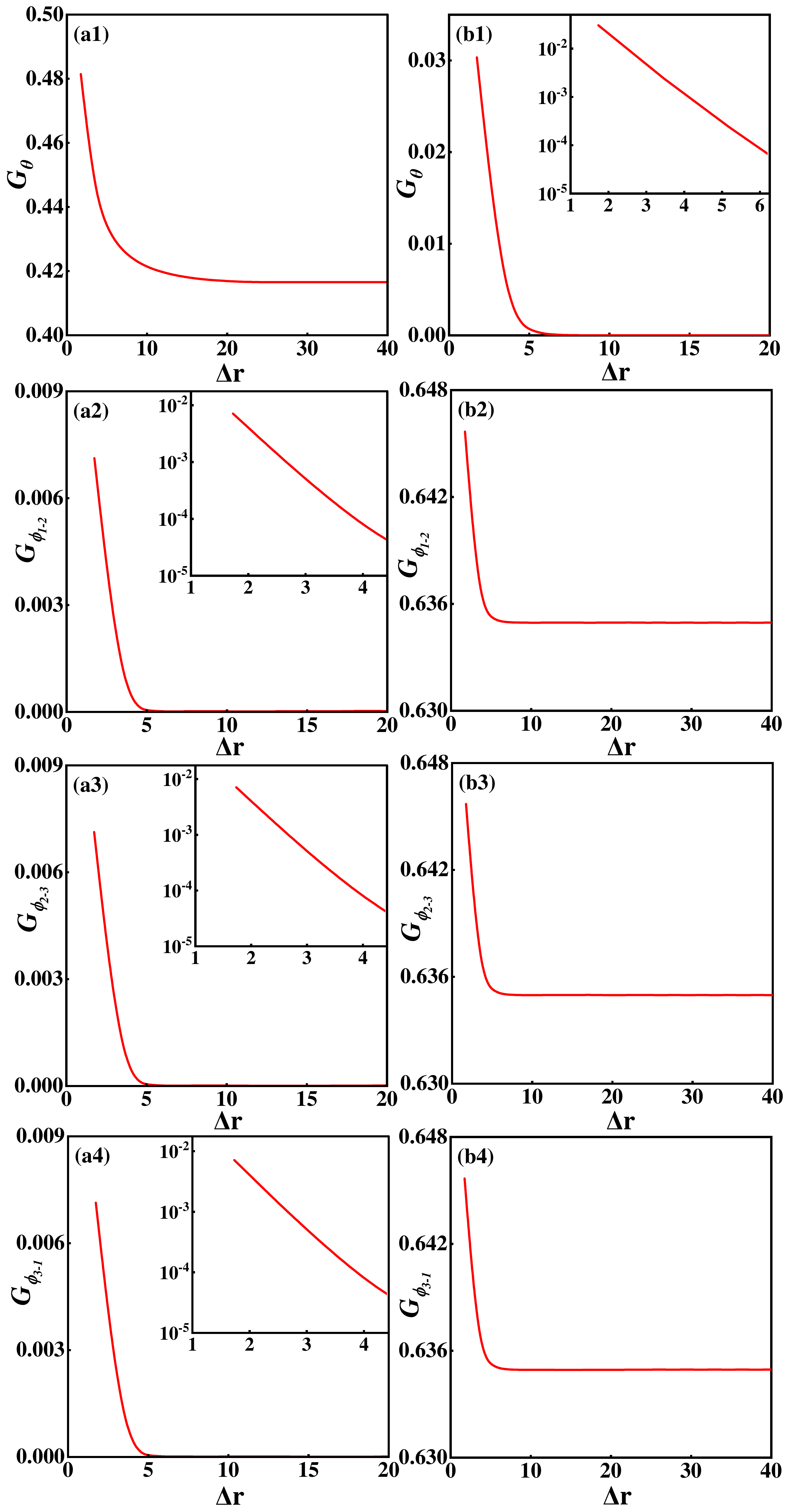}
	\caption{(Color online) The correlation function $G_{\theta/\phi_{a-b}}$ (the $T_{2g}/T_{1u}$ representations with $A=0.16\rho$) for (a1)-(a4) for the point $\bm{A}$ ($\kappa=0.1\rho, T=0.2\rho$) for (b1)-(b4) for the point $\bm{D}$ ($\kappa=0.4\rho, T=0.35\rho$) labeled in Fig.~\ref{T2g_chiral}. Insets of (a2)-(a4) and (b1) only the y-axis is logarithmic.}\label{T2g_chiral_all_AD}
\end{figure}

\begin{figure}[t!]
	\centering
	\includegraphics[width=0.4\textwidth]{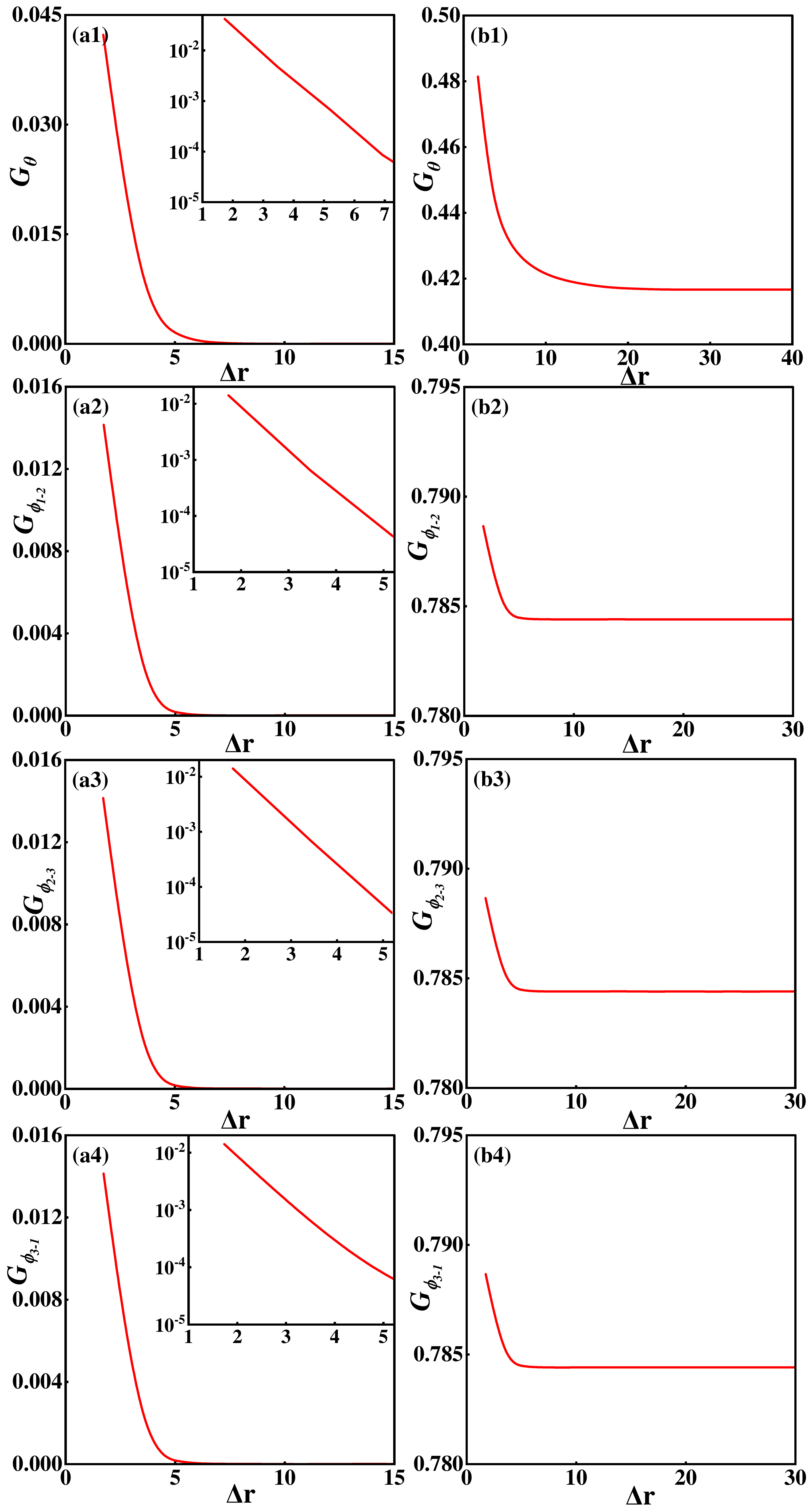}
	\caption{(Color online) The correlation function $G_{\theta/\phi_{a-b}}$ (the $T_{2g}/T_{1u}$ representations with $A=0.16\rho$) for (a1)-(a4) for the point $\bm{B}$ ($\kappa=0.2\rho, T=0.32\rho$), for (b1)-(b4) for the point $\bm{C}$ ($\kappa=0.3\rho, T=0.2\rho$) labeled in Fig.~\ref{T2g_chiral}. Insets of (a1)-(a4) only the y-axis is logarithmic.}\label{T2g_chiral_all_BC}
\end{figure}

In Fig.~\ref{T2g_chiral_ob_2}, we show the thermodynamic quantities as functions of temperature for lattice size $L=30, 35, 40$ at $\kappa/\rho=0.15,0.21$ and $0.3$. More detailedly, Fig.~\ref{T2g_chiral_ob_2}(a1-a3), (b1-b3) and (c1-c3) show the susceptibility $\chi_{\phi_{1-2}}$ of $\phi_1-\phi_2$, the susceptibility $\chi_{\phi_{2-3}}$ of $\phi_2-\phi_3$ and the susceptibility $\chi_{\phi_{3-1}}$ of $\phi_3-\phi_1$, respectively. Fig.~\ref{T2g_chiral_ob_2}(d1-d3), (e1-e3) and (f1-f3) show $\frac{3}{2}U_{\phi_{1-2}}$, where $U_{\phi_{1-2}}$ is the Binder cumulant of the $\phi_1-\phi_2$ field, $\frac{3}{2}U_{\phi_{2-3}}$, where $U_{\phi_{2-3}}$ is the Binder cumulant of the $\phi_2-\phi_3$ field and $\frac{3}{2}U_{\phi_{3-1}}$, where $U_{\phi_{3-1}}$ is the Binder cumulant of the $\phi_3-\phi_1$ field, respectively. Fig.~\ref{T2g_chiral_ob_2}(g1-g3), (h1-h3) and (i1-i3) show the Ising order parameter $I_{1-2}$ of $\phi_1-\phi_2$, the Ising order parameter $I_{2-3}$ of $\phi_2-\phi_3$ and the Ising order parameter $I_{3-1}$ of $\phi_3-\phi_1$, respectively. The grey dotted lines in (a1-i3) mark the phase transitions. Our results demonstrate that the three susceptibilities $\chi_{\phi_{a-b}}$, the three cumulants $\frac{3}{2}U_{\phi_{a-b}}$ and the three Ising order parameters $I_{a-b}$ all exhibit consistent behavior.

Moreover, the spatial correlation functions $G_{\theta}$ and $G_{\phi_{a-b}} $ are shown in Fig.~\ref{T2g_chiral_all_AD}(a1)-(a4) for the typical point $\bm{A}$ marked in Fig.~\ref{T2g_chiral}: while $G_{\theta}$ saturates to a finite value as $\Delta r \to \infty$, $G_{\phi_{a-b}}$ decays exponentially with $\Delta r$, consistent with the charge-$6e$ SC. Fig.~\ref{T2g_chiral_all_AD}(b1)-(b4) are for the typical point $\bm{D}$ marked in Fig.~\ref{T2g_chiral}: while $G_{\theta}$ decays exponentially with $\Delta r$, $G_{\phi_{a-b}}$ saturates to a finite value as $\Delta r \to \infty$, consistent with the chiral metal. Fig.~\ref{T2g_chiral_all_BC}(a1)-(a4) are for the typical point $\bm{B}$ marked in Fig.~\ref{T2g_chiral}: both $G_{\theta}$ and $G_{\phi_{a-b}}$ decay exponentially with $\Delta r$, consistent with the metal. Fig.~\ref{T2g_chiral_all_BC}(b1)-(b4) are for the typical point $\bm{C}$ marked in Fig.~\ref{T2g_chiral}: both $G_{\theta}$ and $G_{\phi_{a-b}}$ saturate to a finite value as $\Delta r \to \infty$, consistent with the chiral SC.

\section{Detailed Results for MC study of $T_{2g}/T_{1u}$ chiral SC with $A=0$}

For the chiral SC within three-dimensional IRRPs $T_{2g}/T_{1u}$ with $A=0$, the $\phi_a - \phi_b$ fields undergo a phase transition of the XY universality class. The susceptibility $\chi$ and the Binder cumulant $U$ of the $\phi_{a}-\phi_{b}$ fields are given as~\cite{janke1990}
\begin{eqnarray}
\chi=\dfrac{N(\left\langle m^2\right\rangle-\left\langle |m|\right\rangle^2)}{K_BT},~~~~~~U=1-\dfrac{\left\langle m^4\right\rangle}{3\left\langle m^2\right\rangle^2},
\end{eqnarray}
where $m_{\phi_{a-b}}=\frac{1}{N}\sum_i e^{i(\theta_a(\bm{r}_{i})-\theta_b(\bm{r}_{i}))}$. The $3U_{\phi_{a-b}}-1$ for the XY-type transition in the $\phi_a-\phi_b$ field.

\begin{figure}[t!]
	\centering
	\includegraphics[width=0.9\linewidth]{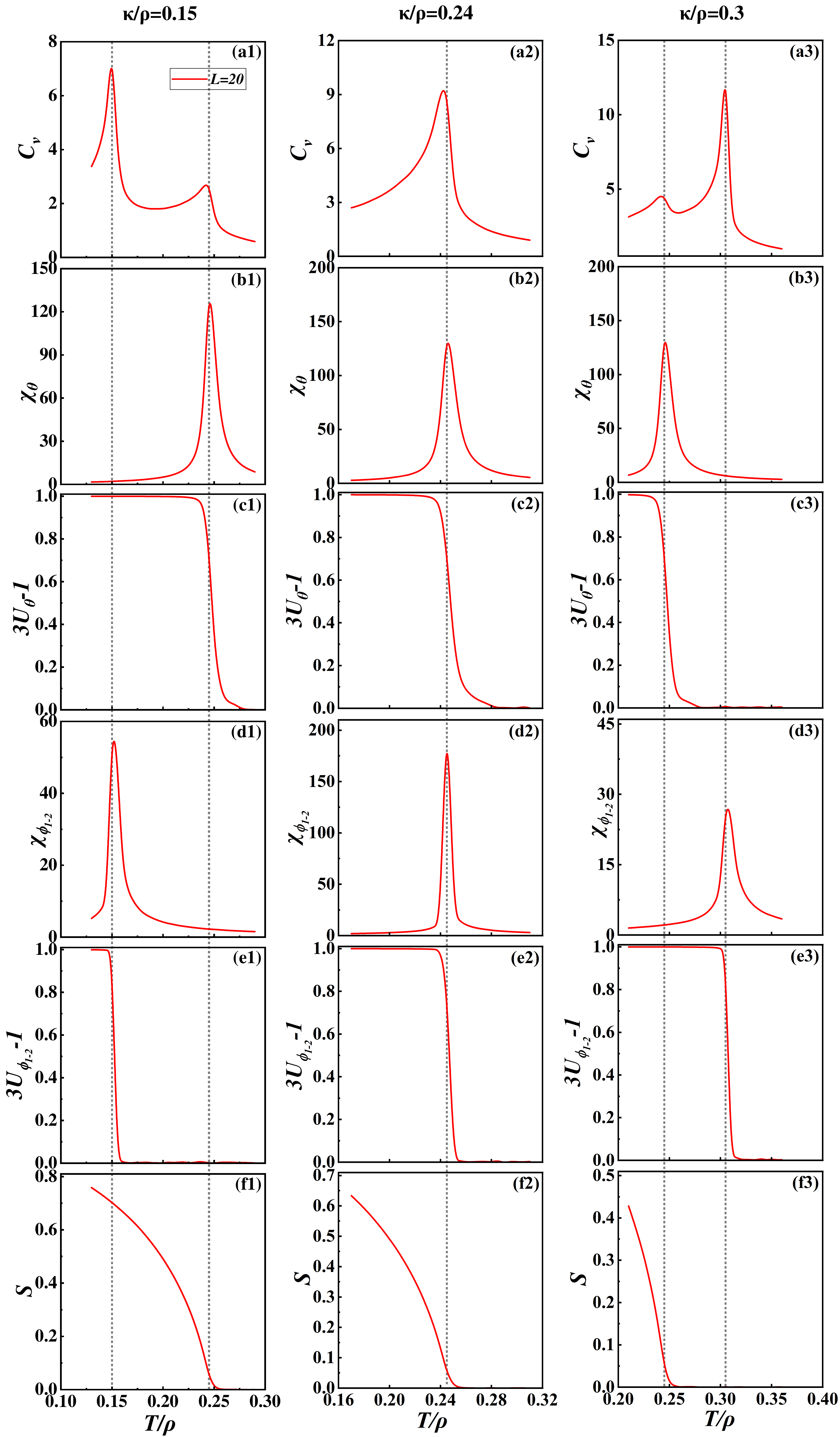}
	\caption{(Color online) The quantities as functions of temperature for $\kappa/\rho=0.15$ (a1,b1,...,f1), $\kappa/\rho=0.24$ (a2,b2,...,f2) and $\kappa/\rho=0.3$ (a3,b3,...,f3) (the $T_{2g}/T_{1u}$ representations with $A=0$). The scaling in all figures is $L=$ 20 (red line). (a1-a3) The specific heat $C_v$. (b1-b3) The susceptibility $\chi_{\theta}$ of $\theta$ field. (c1-c3) $3U_{\theta}-1$, where $U_{\theta}$ is the Binder cumulant of the $\theta$-field. (d1-d3) The susceptibility $\chi_{\phi_{1-2}}$ of $\phi_1-\phi_2$ field. (e1-e3) $3U_{\phi_{1-2}}-1$, where $U_{\phi_{1-2}}$ is the Binder cumulant of the $\phi_1-\phi_2$ field. (f1-f3) The phase stiffness $S$ of $\theta$ field. The grey dashed lines represent the phase transitions in (a1)-(f3).}\label{T2g_chiral_ob_3}
\end{figure}

Thermodynamic data in Fig.~\ref{T2g_chiral_ob_3} indicate three distinct transition sequences depending on the stiffness ratio $\kappa/\rho$:

(i) Low stiffness ratio ($\kappa/\rho=0.15$): As shown in panels (a1,b1,...,f1) of Fig.~\ref{T2g_chiral_ob_3}, two well-separated transitions occur upon heating. The first transition at $T/\rho \approx 0.15$ is characterized by a sharp peak in $C_v$ and a divergence in $\chi_{\phi_{1-2}}$, signaling the disordering of the relative phase $\phi_1$-$\phi_2$. The vanishing of the cumulant $3U_{\phi_{1-2}}-1$ confirms this transition belongs to the Ising universality class. With the $\theta$ field remaining ordered, the system enters a vestigial charge-$6e$ superconducting phase. A second transition appears at $T/\rho \approx 0.245$, marked by another $C_v$ peak. Here, the susceptibility $\chi_{\theta}$ of the global phase diverges while $3U_{\theta}-1$ vanishes, indicating the loss of $\theta$-field coherence and the transition into the normal metal phase.

(ii) Intermediate stiffness ratio ($\kappa/\rho=0.24$): A single simultaneous transition is observed at $T/\rho \approx 0.245$ (panels (a2,b2,...,f2)). This is evidenced by a sharp peak in $C_v$ together with the concurrent vanishing of the phase stiffness $S$ and $3U_{\theta}-1$. Simultaneously, $\chi_{\theta}$ and $\chi_{\phi_{1-2}}$ diverge, indicating that the global phase $\theta$ and the relative phases $\phi_1-\phi_2$ disorder at the same temperature. The system thus undergoes a direct transition from the chiral superconducting phase to the normal metal phase.

(iii) High stiffness ratio ($\kappa/\rho=0.3$): The transition sequence is reversed compared to the $\kappa/\rho=0.15$ case (panels (a3,b3,...,f3)). The first transition occurs at $T/\rho \approx 0.245$, signaled by a peak in $C_v$. At this point, the $\theta$-field disorders, as indicated by $\chi_{\theta}$ becoming divergent and the vanishing of the phase stiffness $S$, driving the system into a chiral metal phase where the relative phase order persists. A subsequent transition takes place at $T/\rho \approx 0.305$, marked by another $C_v$ peak together with diverging $\chi_{\phi_{1-2}}$ and a vanishing $3U_{\phi_{1-2}}-1$. This corresponds to the disordering of the relative phases via an XY transition, finally leading to the normal metal phase.

\begin{figure}[t!]
	\centering
	\includegraphics[width=0.9\linewidth]{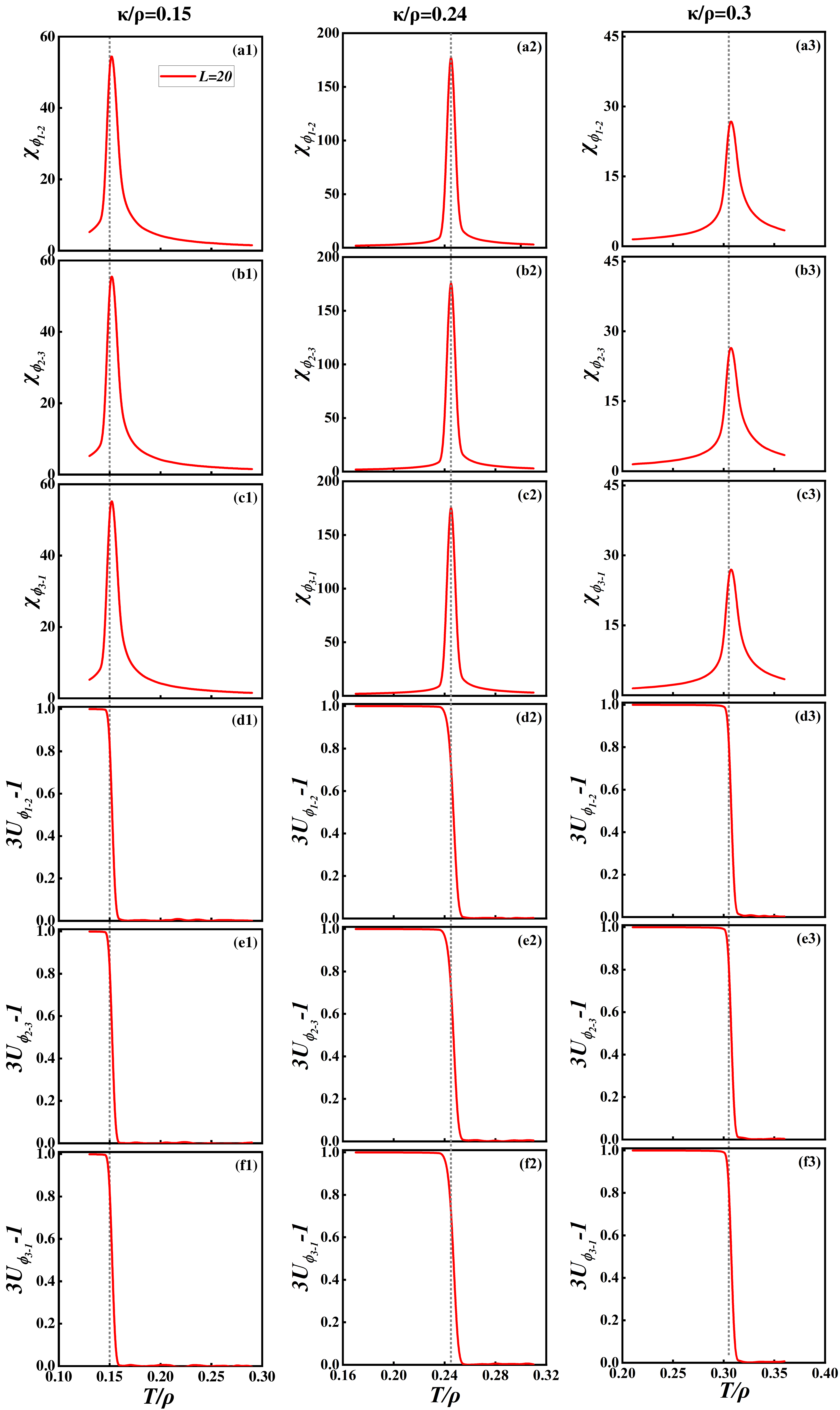}
	\caption{(Color online) The quantities as functions of temperature for $\kappa/\rho=0.15$ (a1,b1,..,f1), $\kappa/\rho=0.24$ (a2,b2,...,f2) and $\kappa/\rho=0.3$ (a3,b3,...,f3) (the $T_{2g}/T_{1u}$ representations with $A=0$). The scaling in all figures is $L=$ 20 (red line). (a1-a3), (b1-b3) and (c1-c3) The susceptibility $\chi_{\phi_{a-b}}$ of $\phi_a-\phi_b$ field, respectively. (d1-d3), (e1-e3) and (f1-f3) $3U_{\phi_{a-b}}-1$, where $U_{\phi_{a-b}}$ is the Binder cumulant of the $\phi_a-\phi_b$ field, respectively. The grey dashed lines represent the phase transitions in (a1)-(f3).}\label{T2g_chiral_ob_4}
\end{figure}

In Fig.~\ref{T2g_chiral_ob_4}, the thermodynamic quantities are plotted against temperature for a lattice size of $L=20$ at $\kappa/\rho=0.15, 0.24$, and $0.3$. Specifically, panels (a1–a3), (b1–b3), and (c1–c3) display the susceptibility $\chi_{\phi_{1-2}}$ of the $\phi_1-\phi_2$ field, $\chi_{\phi_{2-3}}$ of the $\phi_2-\phi_3$ field, and $\chi_{\phi_{3-1}}$ of the $\phi_3-\phi_1$ field, respectively. Correspondingly, panels (d1–d3), (e1–e3), and (f1–f3) show the quantities $3U_{\phi_{1-2}}-1$, $3U_{\phi_{2-3}}-1$, and $3U_{\phi_{3-1}}-1$, where $U_{\phi_{a-b}}$ denotes the Binder cumulant for the respective $\phi_{a-b}$ field. Phase transitions are indicated by grey dotted lines in all subfigures (a1–f3). The results reveal that the three susceptibilities $\chi_{\phi_{a-b}}$ and the three rescaled cumulants $3U_{\phi_{a-b}}-1$ all exhibit consistent behavior across the transitions.

%\end{widetext}

\bibliography{references}

@Article{agterberg2008dis,
author={Agterberg, D. F.
and Tsunetsugu, H.},
title={Dislocations and vortices in pair-density-wave superconductors},
journal={Nat. Phys.},
year={2008},
month={Aug},
day={01},
volume={4},
number={8},
pages={639-642},
issn={1745-2481},
doi={10.1038/nphys999},
url={https://doi.org/10.1038/nphys999}
}

@Article{berg2009charge4e,
author={Berg, Erez
and Fradkin, Eduardo
and Kivelson, Steven A.},
title={Charge-4e superconductivity from pair-density-wave order in certain high-temperature superconductors},
journal={Nat. Phys.},
year={2009},
month={Nov},
day={01},
volume={5},
number={11},
pages={830-833},
issn={1745-2481},
doi={10.1038/nphys1389},
url={https://doi.org/10.1038/nphys1389}
}

@article{agterberg2011con,
  title = {Conventional and charge-six superfluids from melting hexagonal Fulde-Ferrell-Larkin-Ovchinnikov phases in two dimensions},
  author = {Agterberg, D. F. and Geracie, M. and Tsunetsugu, H.},
  journal = {Phys. Rev. B},
  volume = {84},
  issue = {1},
  pages = {014513},
  numpages = {7},
  year = {2011},
  month = {Jul},
  publisher = {American Physical Society},
  doi = {10.1103/PhysRevB.84.014513},
  url = {https://link.aps.org/doi/10.1103/PhysRevB.84.014513}
}

@article{babaev2004phase,
title = {Phase diagram of planar \text{U(1)×U(1)} superconductor: Condensation of vortices with fractional flux and a superfluid state},
journal = {Nucl. Phys. B},
volume = {686},
number = {3},
pages = {397-412},
year = {2004},
issn = {0550-3213},
doi = {https://doi.org/10.1016/j.nuclphysb.2004.02.021},
url = {https://www.sciencedirect.com/science/article/pii/S0550321304001294},
author = {Egor Babaev},
}

@article{ko2009doped,
  title = {Doped kagome system as exotic superconductor},
  author = {Ko, Wing-Ho and Lee, Patrick A. and Wen, Xiao-Gang},
  journal = {Phys. Rev. B},
  volume = {79},
  issue = {21},
  pages = {214502},
  numpages = {13},
  year = {2009},
  month = {Jun},
  publisher = {American Physical Society},
  doi = {10.1103/PhysRevB.79.214502},
  url = {https://link.aps.org/doi/10.1103/PhysRevB.79.214502}
}

@article{herland2010phase,
  title = {Phase transitions in a three dimensional \text{U(1)×U(1)} lattice London superconductor: Metallic superfluid and charge-$4e$ superconducting states},
  author = {Herland, Egil V. and Babaev, Egor and Sudb\o{}, Asle},
  journal = {Phys. Rev. B},
  volume = {82},
  issue = {13},
  pages = {134511},
  numpages = {16},
  year = {2010},
  month = {Oct},
  publisher = {American Physical Society},
  doi = {10.1103/PhysRevB.82.134511},
  url = {https://link.aps.org/doi/10.1103/PhysRevB.82.134511}
}

@article{song2022phase,
  title = {Phase Coherence of Pairs of Cooper Pairs as Quasi-Long-Range Order of Half-Vortex Pairs in a Two-Dimensional Bilayer System},
  author = {Song, Feng-Feng and Zhang, Guang-Ming},
  journal = {Phys. Rev. Lett.},
  volume = {128},
  issue = {19},
  pages = {195301},
  numpages = {6},
  year = {2022},
  month = {May},
  publisher = {American Physical Society},
  doi = {10.1103/PhysRevLett.128.195301},
  url = {https://link.aps.org/doi/10.1103/PhysRevLett.128.195301}
}

@article{li2024charge4e,
title = {Charge-$4e$ superconductor: A wavefunction approach},
journal = {Sci. Bull.},
volume = {69},
number = {15},
pages = {2328-2331},
year = {2024},
issn = {2095-9273},
doi = {https://doi.org/10.1016/j.scib.2024.06.002},
url = {https://www.sciencedirect.com/science/article/pii/S2095927324003980},
author = {Pengfei Li and Kun Jiang and Jiangping Hu}
}

@article{zhang2024higgs,
author={Zhang, Ling-Feng
and Wang, Zhi
and Hu, Xiao},
title={Higgs-Leggett mechanism for the elusive $\phi_0/3=hc/6e$ oscillation in Little-Parks setup of Kagome superconductor {C}s{V}$_3${S}b$_5$},
journal={Commun. Phys.},
year={2024},
month={Jun},
day={29},
volume={7},
number={1},
pages={210},
issn={2399-3650},
doi={10.1038/s42005-024-01663-0},
url={https://doi.org/10.1038/s42005-024-01663-0}
}

@article{zhou2022chern,
author={Zhou, Sen
and Wang, Ziqiang},
title={Chern Fermi pocket, topological pair density wave, and charge-4e and charge-6e superconductivity in kagom{\'e} superconductors},
journal={Nat. Commun.},
year={2022},
month={Nov},
day={26},
volume={13},
number={1},
pages={7288},
issn={2041-1723},
doi={10.1038/s41467-022-34832-2},
url={https://doi.org/10.1038/s41467-022-34832-2}
}

@article{rampp2022topo,
  title = {Topologically Enabled Superconductivity},
  author = {Rampp, Michael A. and K\"onig, Elio J. and Schmalian, J\"org},
  journal = {Phys. Rev. Lett.},
  volume = {129},
  issue = {7},
  pages = {077001},
  numpages = {6},
  year = {2022},
  month = {Aug},
  publisher = {American Physical Society},
  doi = {10.1103/PhysRevLett.129.077001},
  url = {https://link.aps.org/doi/10.1103/PhysRevLett.129.077001}
}

@article{yu2023non,
  title = {Nondegenerate surface pair density wave in the kagome superconductor  $\text{Cs}$$\text{V}_{3}$$\text{Sb}_{5}$: Application to vestigial orders},
  author = {Yu, Yue},
  journal = {Phys. Rev. B},
  volume = {108},
  issue = {5},
  pages = {054517},
  numpages = {7},
  year = {2023},
  month = {Aug},
  publisher = {American Physical Society},
  doi = {10.1103/PhysRevB.108.054517},
  url = {https://link.aps.org/doi/10.1103/PhysRevB.108.054517}
}

@article{curtis2023stabliz,
  title = {Stabilizing Fluctuating Spin-Triplet Superconductivity in Graphene via Induced Spin-Orbit Coupling},
  author = {Curtis, Jonathan B. and Poniatowski, Nicholas R. and Xie, Yonglong and Yacoby, Amir and Demler, Eugene and Narang, Prineha},
  journal = {Phys. Rev. Lett.},
  volume = {130},
  issue = {19},
  pages = {196001},
  numpages = {7},
  year = {2023},
  month = {May},
  publisher = {American Physical Society},
  doi = {10.1103/PhysRevLett.130.196001},
  url = {https://link.aps.org/doi/10.1103/PhysRevLett.130.196001}
}

@Article{poduval2024vestigial,
author={Poduval, Prathyush P.
and Scheurer, Mathias S.},
title={Vestigial singlet pairing in a fluctuating magnetic triplet superconductor and its implications for graphene superlattices},
journal={Nat. Commun.},
year={2024},
month={Feb},
day={24},
volume={15},
number={1},
pages={1713},
issn={2041-1723},
doi={10.1038/s41467-024-45950-4},
url={https://doi.org/10.1038/s41467-024-45950-4}
}

@Article{zeng2024high,
author={Zeng, Meng
and Hu, Lun-Hui
and Hu, Hong-Ye
and You, Yi-Zhuang
and Wu, Congjun},
title={High-order time-reversal symmetry breaking normal state},
journal={Sci. China- Phys. Mech. Astron.},
year={2024},
month={Feb},
day={02},
volume={67},
number={3},
pages={237411},
issn={1869-1927},
doi={10.1007/s11433-023-2287-8},
url={https://doi.org/10.1007/s11433-023-2287-8}
}

@article{hecker2023cascade,
  title = {Cascade of vestigial orders in two-component superconductors: Nematic, ferromagnetic, $s$-wave charge-$4e$, and $d$-wave charge-$4e$ states},
  author = {Hecker, Matthias and Willa, Roland and Schmalian, J\"org and Fernandes, Rafael M.},
  journal = {Phys. Rev. B},
  volume = {107},
  issue = {22},
  pages = {224503},
  numpages = {26},
  year = {2023},
  month = {Jun},
  publisher = {American Physical Society},
  doi = {10.1103/PhysRevB.107.224503},
  url = {https://link.aps.org/doi/10.1103/PhysRevB.107.224503}
}

@article{hecker2024local,
  title = {Local condensation of charge-$4e$ superconductivity at a nematic domain wall},
  author = {Hecker, Matthias and Fernandes, Rafael M.},
  journal = {Phys. Rev. B},
  volume = {109},
  issue = {13},
  pages = {134514},
  numpages = {8},
  year = {2024},
  month = {Apr},
  publisher = {American Physical Society},
  doi = {10.1103/PhysRevB.109.134514},
  url = {https://link.aps.org/doi/10.1103/PhysRevB.109.134514}
}

@article{jian2021charge4e,
  title = {Charge-$4e$ Superconductivity from Nematic Superconductors in Two and Three Dimensions},
  author = {Jian, Shao-Kai and Huang, Yingyi and Yao, Hong},
  journal = {Phys. Rev. Lett.},
  volume = {127},
  issue = {22},
  pages = {227001},
  numpages = {6},
  year = {2021},
  month = {Nov},
  publisher = {American Physical Society},
  doi = {10.1103/PhysRevLett.127.227001},
  url = {https://link.aps.org/doi/10.1103/PhysRevLett.127.227001}
}

@article{fu2021charge4e,
  title = {Charge-$4e$ Superconductivity from Multicomponent Nematic Pairing: Application to Twisted Bilayer Graphene},
  author = {Fernandes, Rafael M. and Fu, Liang},
  journal = {Phys. Rev. Lett.},
  volume = {127},
  issue = {4},
  pages = {047001},
  numpages = {6},
  year = {2021},
  month = {Jul},
  publisher = {American Physical Society},
  doi = {10.1103/PhysRevLett.127.047001},
  url = {https://link.aps.org/doi/10.1103/PhysRevLett.127.047001}
}

@article{grinenko2021state,
author={Grinenko, Vadim and Weston, Daniel and Caglieris, Federico and Wuttke, Christoph and Hess, Christian and Gottschall, Tino and Maccari, Ilaria and Gorbunov, Denis and Zherlitsyn, Sergei and Wosnitza, Jochen and Rydh, Andreas and Kihou, Kunihiro and Lee, Chul-Ho and Sarkar, Rajib and Dengre, Shanu and Garaud, Julien and Charnukha, Aliaksei and H{\"u}hne, Ruben and Nielsch, Kornelius and B{\"u}chner, Bernd and Klauss, Hans-Henning and Babaev, Egor},
title={State with spontaneously broken time-reversal symmetry above the superconducting phase transition},
journal={Nat. Phys.},
year={2021},
month={Nov},
day={01},
volume={17},
number={11},
pages={1254-1259},
issn={1745-2481},
url={https://doi.org/10.1038/s41567-021-01350-9}
}

@Article{wu2024dwave,
author={Wu, Yi-Ming and Wang, Yuxuan},
title={$d$-wave charge-$4e$ superconductivity from fluctuating pair density waves},
journal={npj Quantum Mater.},
year={2024},
month={Sep},
day={05},
volume={9},
number={1},
pages={66},
issn={2397-4648},
url={https://doi.org/10.1038/s41535-024-00674-y}
}

@article{you2012super,
  title = {Superfluidity of Bosons in Kagome Lattices with Frustration},
  author = {You, Yi-Zhuang and Chen, Zhu and Sun, Xiao-Qi and Zhai, Hui},
  journal = {Phys. Rev. Lett.},
  volume = {109},
  issue = {26},
  pages = {265302},
  numpages = {5},
  year = {2012},
  month = {Dec},
  publisher = {American Physical Society},
  doi = {10.1103/PhysRevLett.109.265302},
  url = {https://link.aps.org/doi/10.1103/PhysRevLett.109.265302}
}

@article{korshunov1985,
  title={Two-dimensional superfluid Fermi liquid with $p$-pairing},
  author={Korshunov, SE},
  journal={Sov. Phys. JETP},
  volume={62},
  issue = {2},
  pages={301},
  year = {1985},
  url = {http://www.jetp.ras.ru/cgi-bin/dn/e_062_02_0301.pdf}
}

@article{kivelson1990doped,
  title = {Doped antiferromagnets in the weak-hopping limit},
  author = {Kivelson, S. A. and Emery, V. J. and Lin, H. Q.},
  journal = {Phys. Rev. B},
  volume = {42},
  issue = {10},
  pages = {6523--6530},
  numpages = {0},
  year = {1990},
  month = {Oct},
  publisher = {American Physical Society},
  doi = {10.1103/PhysRevB.42.6523},
  url = {https://link.aps.org/doi/10.1103/PhysRevB.42.6523}
}

@article{ropke1998four,
  title = {Four-Particle Condensate in Strongly Coupled Fermion Systems},
  author = {R\"opke, G. and Schnell, A. and Schuck, P. and Nozi\`eres, P.},
  journal = {Phys. Rev. Lett.},
  volume = {80},
  issue = {15},
  pages = {3177--3180},
  numpages = {0},
  year = {1998},
  month = {Apr},
  publisher = {American Physical Society},
  doi = {10.1103/PhysRevLett.80.3177},
  url = {https://link.aps.org/doi/10.1103/PhysRevLett.80.3177}
}

@article{douccot2002pairing,
  title={Pairing of Cooper pairs in a fully frustrated Josephson-junction chain},
  author={Dou{\c{c}}ot, Benoit and Vidal, Julien},
  journal = {Phys. Rev. Lett.},
  volume={88},
  number={22},
  pages={227005},
  numpages = {4},
  year={2002},
  publisher = {American Physical Society},
  doi = {10.1103/PhysRevLett.88.227005},
  url = {https://link.aps.org/doi/10.1103/PhysRevLett.88.227005}
}

@article{moore2004,
  title = {Geometric effects on T-breaking in $p+ip$ and $d+id$ superconducting arrays},
  author = {Moore, J. E. and Lee, D.-H.},
  journal = {Phys. Rev. B},
  volume = {69},
  issue = {10},
  pages = {104511},
  numpages = {8},
  year = {2004},
  month = {Mar},
  publisher = {American Physical Society},
  doi = {10.1103/PhysRevB.69.104511},
  url = {https://link.aps.org/doi/10.1103/PhysRevB.69.104511}
}

@article{wu2005competing,
  title = {Competing Orders in One-Dimensional Spin-$3/2$ Fermionic Systems},
  author = {Wu, Congjun},
  journal = {Phys. Rev. Lett.},
  volume = {95},
  issue = {26},
  pages = {266404},
  numpages = {4},
  year = {2005},
  month = {Dec},
  publisher = {American Physical Society},
  doi = {10.1103/PhysRevLett.95.266404},
  url = {https://link.aps.org/doi/10.1103/PhysRevLett.95.266404}
}

@article{aligia2005quartet,
  title = {Quartet Formation at $(100)/(110)$ Interfaces of $d$-Wave Superconductors},
  author = {Aligia, A. A. and Kampf, A. P. and Mannhart, J.},
  journal = {Phys. Rev. Lett.},
  volume = {94},
  issue = {24},
  pages = {247004},
  numpages = {4},
  year = {2005},
  month = {Jun},
  publisher = {American Physical Society},
  doi = {10.1103/PhysRevLett.94.247004},
  url = {https://link.aps.org/doi/10.1103/PhysRevLett.94.247004}
}

@article{jiang2017charge4e,
  title = {Charge-$4e$ superconductors: A Majorana quantum Monte Carlo study},
  author = {Jiang, Yi-Fan and Li, Zi-Xiang and Kivelson, Steven A. and Yao, Hong},
  journal = {Phys. Rev. B},
  volume = {95},
  issue = {24},
  pages = {241103},
  numpages = {5},
  year = {2017},
  month = {Jun},
  publisher = {American Physical Society},
  doi = {10.1103/PhysRevB.95.241103},
  url = {https://link.aps.org/doi/10.1103/PhysRevB.95.241103}
}

@article{ge2024charge4e,
  title = {Charge-$4e$ and Charge-$6e$ Flux Quantization and Higher Charge Superconductivity in Kagome Superconductor Ring Devices},
  author = {Ge, Jun and Wang, Pinyuan and Xing, Ying and Yin, Qiangwei and Wang, Anqi and Shen, Jie and Lei, Hechang and Wang, Ziqiang and Wang, Jian},
  journal = {Phys. Rev. X},
  volume = {14},
  issue = {2},
  pages = {021025},
  numpages = {19},
  year = {2024},
  month = {May},
  publisher = {American Physical Society},
  doi = {10.1103/PhysRevX.14.021025},
  url = {https://link.aps.org/doi/10.1103/PhysRevX.14.021025}
}

@article{han2022,
  title = {Understanding resistance oscillation in the $\text{Cs}$$\text{V}_{3}$$\text{Sb}_{5}$ superconductor},
  author = {Han, Jung Hoon and Lee, Patrick A.},
  journal = {Phys. Rev. B},
  volume = {106},
  issue = {18},
  pages = {184515},
  numpages = {8},
  year = {2022},
  month = {Nov},
  publisher = {American Physical Society},
  doi = {10.1103/PhysRevB.106.184515},
  url = {https://link.aps.org/doi/10.1103/PhysRevB.106.184515}
}

@article{li2017nematic,
author={Li, Jun
and Pereira, Paulo J.
and Yuan, Jie
and Lv, Yang-Yang
and Jiang, Mei-Ping
and Lu, Dachuan
and Lin, Zi-Quan
and Liu, Yong-Jie
and Wang, Jun-Feng
and Li, Liang
and Ke, Xiaoxing
and Van Tendeloo, Gustaaf
and Li, Meng-Yue
and Feng, Hai-Luke
and Hatano, Takeshi
and Wang, Hua-Bing
and Wu, Pei-Heng
and Yamaura, Kazunari
and Takayama-Muromachi, Eiji
and Vanacken, Johan
and Chibotaru, Liviu F.
and Moshchalkov, Victor V.},
title={Nematic superconducting state in iron pnictide superconductors},
journal={Nat. Commun.},
year={2017},
month={Dec},
day={01},
volume={8},
number={1},
pages={1880},
issn={2041-1723},
doi={10.1038/s41467-017-02016-y},
url={https://doi.org/10.1038/s41467-017-02016-y}
}

@Article{yonezawa2017therm,
author={Yonezawa, Shingo
and Tajiri, Kengo
and Nakata, Suguru
and Nagai, Yuki
and Wang, Zhiwei
and Segawa, Kouji
and Ando, Yoichi
and Maeno, Yoshiteru},
title={Thermodynamic evidence for nematic superconductivity in $\text{Cu}_x$$\text{Bi}_{2}$$\text{Se}_{3}$},
journal={Nat. Phys.},
year={2017},
month={Feb},
day={01},
volume={13},
number={2},
pages={123-126},
issn={1745-2481},
doi={10.1038/nphys3907},
url={https://doi.org/10.1038/nphys3907}
}

@article{tao2018direct,
  title = {Direct Visualization of the Nematic Superconductivity in $\text{Cu}_x$$\text{Bi}_{2}$$\text{Se}_{3}$},
  author = {Tao, Ran and Yan, Ya-Jun and Liu, Xi and Wang, Zhi-Wei and Ando, Yoichi and Wang, Qiang-Hua and Zhang, Tong and Feng, Dong-Lai},
  journal = {Phys. Rev. X},
  volume = {8},
  issue = {4},
  pages = {041024},
  numpages = {9},
  year = {2018},
  month = {Nov},
  publisher = {American Physical Society},
  doi = {10.1103/PhysRevX.8.041024},
  url = {https://link.aps.org/doi/10.1103/PhysRevX.8.041024}
}

@Article{kostylev2020uniaxial,
author={Kostylev, Ivan
and Yonezawa, Shingo
and Wang, Zhiwei
and Ando, Yoichi
and Maeno, Yoshiteru},
title={Uniaxial-strain control of nematic superconductivity in $\text{Sr}_x$$\text{Bi}_{2}$$\text{Se}_{3}$},
journal={Nat. Commun.},
year={2020},
month={Aug},
day={24},
volume={11},
number={1},
pages={4152},
issn={2041-1723},
doi={10.1038/s41467-020-17913-y},
url={https://doi.org/10.1038/s41467-020-17913-y}
}

@Article{lothman2022nematic,
author={L{\"o}thman, Tomas
and Schmidt, Johann
and Parhizgar, Fariborz
and Black-Schaffer, Annica M.},
title={Nematic superconductivity in magic-angle twisted bilayer graphene from atomistic modeling},
journal={Commun. Phys.},
year={2022},
month={Apr},
day={14},
volume={5},
number={1},
pages={92},
issn={2399-3650},
doi={10.1038/s42005-022-00860-z},
url={https://doi.org/10.1038/s42005-022-00860-z}
}

@article{le2020evidence,
title = {Evidence for nematic superconductivity of topological surface states in $\text{Pb}$$\text{Ta}$$\text{Se}_{2}$},
journal = {Sci. Bull.},
volume = {65},
number = {16},
pages = {1349-1355},
year = {2020},
issn = {2095-9273},
doi = {https://doi.org/10.1016/j.scib.2020.04.039},
url = {https://www.sciencedirect.com/science/article/pii/S2095927320302644},
author = {Tian Le and Yue Sun and Hui-Ke Jin and Liqiang Che and Lichang Yin and Jie Li and Guiming Pang and Chunqiang Xu and Lingxiao Zhao and Shunichiro Kittaka and Toshiro Sakakibara and Kazushige Machida and Raman Sankar and Huiqiu Yuan and Genfu Chen and Xiaofeng Xu and Shiyan Li and Yi Zhou and Xin Lu},
}

@article{morong2021direct,
  title = {Disorder-controlled relaxation in a three-dimensional \text{H}ubbard model quantum simulator},
  author = {Morong, W. and Muleady, S. R. and Kimchi, I. and Xu, W. and Nandkishore, R. M. and Rey, A. M. and DeMarco, B.},
  journal = {Phys. Rev. Res.},
  volume = {3},
  issue = {1},
  pages = {L012009},
  numpages = {6},
  year = {2021},
  month = {Jan},
  publisher = {American Physical Society},
  doi = {10.1103/PhysRevResearch.3.L012009},
  url = {https://link.aps.org/doi/10.1103/PhysRevResearch.3.L012009}
}

@article{iskakov2022phase,
  title = {Phase transitions in partial summation methods: Results from the three-dimensional \text{H}ubbard model},
  author = {Iskakov, Sergei and Gull, Emanuel},
  journal = {Phys. Rev. B},
  volume = {105},
  issue = {4},
  pages = {045109},
  numpages = {12},
  year = {2022},
  month = {Jan},
  publisher = {American Physical Society},
  doi = {10.1103/PhysRevB.105.045109},
  url = {https://link.aps.org/doi/10.1103/PhysRevB.105.045109}
}

@article{lenihan2022evaluating,
  title = {Evaluating Second-Order Phase Transitions with Diagrammatic Monte Carlo: N\'eel Transition in the Doped Three-Dimensional \text{H}ubbard Model},
  author = {Lenihan, Connor and Kim, Aaram J. and \ifmmode \check{S}\else \v{S}\fi{}imkovic, Fedor and Kozik, Evgeny},
  journal = {Phys. Rev. Lett.},
  volume = {129},
  issue = {10},
  pages = {107202},
  numpages = {6},
  year = {2022},
  month = {Aug},
  publisher = {American Physical Society},
  doi = {10.1103/PhysRevLett.129.107202},
  url = {https://link.aps.org/doi/10.1103/PhysRevLett.129.107202}
}

@article{shao2024afm,
author={Shao, Hou-Ji
and Wang, Yu-Xuan
and Zhu, De-Zhi
and Zhu, Yan-Song
and Sun, Hao-Nan
and Chen, Si-Yuan
and Zhang, Chi
and Fan, Zhi-Jie
and Deng, Youjin
and Yao, Xing-Can
and Chen, Yu-Ao
and Pan, Jian-Wei},
title={Antiferromagnetic phase transition in a 3\text{D} fermionic \text{H}ubbard model},
journal={Nature},
year={2024},
month={Aug},
day={01},
volume={632},
number={8024},
pages={267-272},
issn={1476-4687},
doi={10.1038/s41586-024-07689-2},
url={https://doi.org/10.1038/s41586-024-07689-2}
}

@Article{langmann2025univer,
author={Langmann, E.
and Lenells, J.},
title={Universality of Mean-Field Antiferromagnetic Order in an Anisotropic 3\text{D} \text{H}ubbard Model at Half-Filling},
journal={J. Stat. Phys.},
year={2025},
month={Jan},
day={13},
volume={192},
number={1},
pages={10},
issn={1572-9613},
doi={10.1007/s10955-024-03390-w},
url={https://doi.org/10.1007/s10955-024-03390-w}
}

@article{song2025extended,
  title = {Extended Metal-Insulator Crossover with Strong Antiferromagnetic Spin Correlation in Half-Filled \text{3D} \text{H}ubbard Model},
  author = {Song, Yu-Feng and Deng, Youjin and He, Yuan-Yao},
  journal = {Phys. Rev. Lett.},
  volume = {134},
  issue = {1},
  pages = {016503},
  numpages = {7},
  year = {2025},
  month = {Jan},
  publisher = {American Physical Society},
  doi = {10.1103/PhysRevLett.134.016503},
  url = {https://link.aps.org/doi/10.1103/PhysRevLett.134.016503}
}

@article{song2025magnetic,
  title = {Magnetic, thermodynamic, and dynamical properties of the three-dimensional fermionic \text{H}ubbard model: A comprehensive Monte Carlo study},
  author = {Song, Yu-Feng and Deng, Youjin and He, Yuan-Yao},
  journal = {Phys. Rev. B},
  volume = {111},
  issue = {3},
  pages = {035123},
  numpages = {28},
  year = {2025},
  month = {Jan},
  publisher = {American Physical Society},
  doi = {10.1103/PhysRevB.111.035123},
  url = {https://link.aps.org/doi/10.1103/PhysRevB.111.035123}
}

@Article{sun2025boosting,
	title={{Boosting determinant quantum Monte Carlo with submatrix updates: Unveiling the phase diagram of the 3D Hubbard model}},
	author={Fanjie Sun and Xiao Yan Xu},
	journal={SciPost Phys.},
	volume={18},
	pages={055},
	year={2025},
	publisher={SciPost},
	doi={10.21468/SciPostPhys.18.2.055},
	url={https://scipost.org/10.21468/SciPostPhys.18.2.055},
}

@article{pan2024octupolar,
  title={Octupolar Weyl Superconductivity from Electron-electron Interaction}, 
  author={Zhiming Pan and Chen Lu and Fan Yang and Congjun Wu},
  journal={arXiv:2411.06932},
  year={2024},
  url={https://arxiv.org/abs/2411.06932}, 
}

@article{pan2024frustrated,
author={Pan, Zhiming
and Lu, Chen
and Yang, Fan
and Wu, Congjun},
title={Frustrated superconductivity and sextetting order},
journal={Sci. China- Phys. Mech. Astron.},
year={2024},
month={Jul},
day={22},
volume={67},
number={8},
pages={287412},
issn={1869-1927},
doi={10.1007/s11433-024-2396-y},
url={https://doi.org/10.1007/s11433-024-2396-y}
}

@article{lin2025theory,
  title = {Theory of the charge-$6e$ condensed phase in kagome-lattice superconductors},
  author = {Lin, Tong-Yu and Song, Feng-Feng and Zhang, Guang-Ming},
  journal = {Phys. Rev. B},
  volume = {111},
  issue = {5},
  pages = {054508},
  numpages = {16},
  year = {2025},
  month = {Feb},
  publisher = {American Physical Society},
  doi = {10.1103/PhysRevB.111.054508},
  url = {https://link.aps.org/doi/10.1103/PhysRevB.111.054508}
}

@article{varma2023extended,
  title = {Extended superconducting fluctuation region and $6e$ and $4e$ flux quantization in a kagome compound with a normal state of \text{3Q} order},
  author = {Varma, Chandra M. and Wang, Ziqiang},
  journal = {Phys. Rev. B},
  volume = {108},
  issue = {21},
  pages = {214516},
  numpages = {10},
  year = {2023},
  month = {Dec},
  publisher = {American Physical Society},
  doi = {10.1103/PhysRevB.108.214516},
  url = {https://link.aps.org/doi/10.1103/PhysRevB.108.214516}
}

@article{liu2023charge4e,
author={Liu, Yu-Bo
and Zhou, Jing
and Wu, Congjun
and Yang, Fan},
title={Charge-$4e$ superconductivity and chiral metal in 45{\textdegree}-twisted bilayer cuprates and related bilayers},
journal={Nat. Commun.},
year={2023},
month={Dec},
day={01},
volume={14},
number={1},
pages={7926},
issn={2041-1723},
doi={10.1038/s41467-023-43782-2},
url={https://doi.org/10.1038/s41467-023-43782-2}
}

@article{liu2024nematic,
  title = {Nematic Superconductivity and Its Critical Vestigial Phases in the Quasicrystal},
  author = {Liu, Yu-Bo and Zhou, Jing and Yang, Fan},
  journal = {Phys. Rev. Lett.},
  volume = {133},
  issue = {13},
  pages = {136002},
  numpages = {8},
  year = {2024},
  month = {Sep},
  publisher = {American Physical Society},
  doi = {10.1103/PhysRevLett.133.136002},
  url = {https://link.aps.org/doi/10.1103/PhysRevLett.133.136002}
}

@article{hasenbusch2019monte,
  title = {Monte Carlo study of an improved clock model in three dimensions},
  author = {Hasenbusch, Martin},
  journal = {Phys. Rev. B},
  volume = {100},
  issue = {22},
  pages = {224517},
  numpages = {19},
  year = {2019},
  month = {Dec},
  publisher = {American Physical Society},
  doi = {10.1103/PhysRevB.100.224517},
  url = {https://link.aps.org/doi/10.1103/PhysRevB.100.224517}
}

@article{volovik2024fermionic,
  title={Fermionic quartet and vestigial gravity},
  author={Volovik, Grigorii Efimovich},
  journal={JETP Letters},
  volume={119},
  number={4},
  pages={330--334},
  year={2024},
  publisher={Springer},
  url={https://link.springer.com/article/10.1134/S002136402460006X}
}

@article{sigrist1991,
  title={Phenomenological theory of unconventional superconductivity},
  author={Sigrist, Manfred and Ueda, Kazuo},
  journal={Rev. Mod. Phys.},
  volume={63},
  number={2},
  pages={239},
  year={1991},
  publisher={APS},
  url={https://journals.aps.org/rmp/abstract/10.1103/RevModPhys.63.239}
}

@article{raugh2010sc,
  title = {Superconductivity in the repulsive Hubbard model: An asymptotically exact weak-coupling solution},
  author = {Raghu, S. and Kivelson, S. A. and Scalapino, D. J.},
  journal = {Phys. Rev. B},
  volume = {81},
  issue = {22},
  pages = {224505},
  numpages = {17},
  year = {2010},
  month = {Jun},
  publisher = {American Physical Society},
  doi = {10.1103/PhysRevB.81.224505},
  url = {https://link.aps.org/doi/10.1103/PhysRevB.81.224505}
}

@article{janke1990,
  title = {Crossover in the XY model from three to two dimensions},
  author = {Janke, Wolfhard and Matsui, Tetsuo},
  journal = {Phys. Rev. B},
  volume = {42},
  issue = {16},
  pages = {10673--10681},
  numpages = {0},
  year = {1990},
  month = {Dec},
  publisher = {American Physical Society},
  doi = {10.1103/PhysRevB.42.10673},
  url = {https://link.aps.org/doi/10.1103/PhysRevB.42.10673}
}

\end{document}